\documentclass[aps,pre,preprint,superscriptaddress,fleqn]{revtex4-1}  

\usepackage{graphicx}
\usepackage{charter}
\usepackage{droidsans}
\usepackage{setspace}
\usepackage{tabularx}
\usepackage[english]{babel}
\usepackage[T1]{fontenc}
\usepackage[utf8]{inputenc}
\usepackage{microtype}
\usepackage{lmodern}
\usepackage{amsmath, amsfonts}
\usepackage{mathtools}
\usepackage{float}
\usepackage[usenames, dvipsnames]{xcolor}
\usepackage{siunitx}
\usepackage{color}
\usepackage{todonotes}
\usepackage{setspace}

\definecolor{NGreen}{rgb}{0.0,0.53,0.0}
\definecolor{CiteBlue}{RGB}{28, 58, 189}
			
\usepackage[breaklinks = true, 
colorlinks = true, 
linkcolor = CiteBlue, 
urlcolor = CiteBlue, 
citecolor = CiteBlue]
{hyperref}

\newcommand\un[2]{$\text{#1}_\text{#2}$}

\begin{document}

\title{Stochastic switching between multistable oscillation patterns\\ of the Min-system}

\author{Artemij Amiranashvili}
\thanks{These authors contributed equally.}
\author{Nikolas D Schnellb{\"a}cher}
\thanks{These authors contributed equally.}
\author{Ulrich S Schwarz}
\email{schwarz@thphys.uni-heidelberg.de}
\affiliation{Institute for Theoretical Physics, Heidelberg University, Philosophenweg 19, 69120 Heidelberg, Germany}
\affiliation{BioQuant Center for Quantitative Biology, Heidelberg University, Im Neuenheimer Feld 267, 69120 Heidelberg, Germany}

\begin{abstract}
The spatiotemporal oscillation patterns of the proteins MinD and MinE are used
by the bacterium \textit{E. coli} to sense its own geometry. Strikingly, both computer 
simulations and experiments have recently shown that for the same geometry of the reaction volume, 
different oscillation patterns can be stable, with stochastic switching between them. Here we use particle-based 
Brownian dynamics simulations to  predict the relative frequency of different oscillation patterns 
over a large range of three-dimensional compartment geometries, in excellent agreement
with experimental results. Fourier analyses as well as pattern 
recognition algorithms are used to automatically identify the different oscillation patterns 
and the switching rates between them. We also identify novel oscillation patterns in three-dimensional compartments with
membrane-covered walls and identify a linear relation between the bound Min-protein densities and the volume-to-surface ratio.
In general, our work shows how geometry sensing is limited by multistability and stochastic fluctuations.
\end{abstract}

\maketitle

\section{Introduction}

	The bacterial Min-proteins are a well studied example of a pattern-forming protein system that gives rise to rich spatiotemporal oscillations. It was discovered as a spatial regulator in bacterial cell division, where it ensures
	symmetric division by precise localization of the divisome
	to midcell \cite{DeBoer1989}. The dynamic nature of this protein system was
    demonstrated by live cell imaging in \textit{E. coli} bacteria, where these proteins oscillate along the longitudinal axis between the cell poles of the rod-shaped bacterium, forming so-called \textit{polar zones} \cite{Hu1999MolMicrobiology, Raskin1999, Raskin1999a, Hu1999PNAS}.  
	
	Most bacteria use a cytoskeletal structure, a so-called \textit{Z-ring}, for the completion of bacterial cytokinesis \cite{Lutkenhaus1997}. This Z-ring self-assembles from filaments of polymerized FtsZ-proteins, the prokaryotic homolog of the eukaryotic protein tubulin \cite{VanDenEnt2001}, which serve as a scaffold structure for midcell constriction and the eventual septum formation in the midplane. If successful, this process creates two equally sized daughter cells with an identical set of genetic information \cite{Bi1991}.
	A necessary prerequisite for successful symmetric cell division is hence the targeted assembly of FtsZ towards midcell. In \textit{E. coli} cells this is mediated by two independent mechanisms, nucleoid occlusion and the dynamic oscillation of the MinCDE proteins \cite{Wu2011, Hu1999PNAS, Bailey2014}. While nucleoid occlusion prevents division near the chromosome, the Min-system actively keeps the divisome away from the cell poles through the MinC-protein acting as FtsZ-polymerization inhibitor \cite{Hu1999PNAS}. The characteristic pole-to-pole oscillations create a time-averaged concentration gradient with a minimal inhibitor concentration of MinC at midcell, suppressing Z-ring assembly at the cell poles \cite{Raskin1999, Raskin1999a, Hu1999PNAS}. Although MinC is indispensable for correct division site placement, it acts only as a passenger molecule, passively following the oscillatory dynamics of MinD and MinE \cite{Hu1999MolMicrobiology, Raskin1999, DeBoer1992a}.
	
\begin{figure}[t]
	\centering
	\includegraphics[width = 0.95\textwidth]{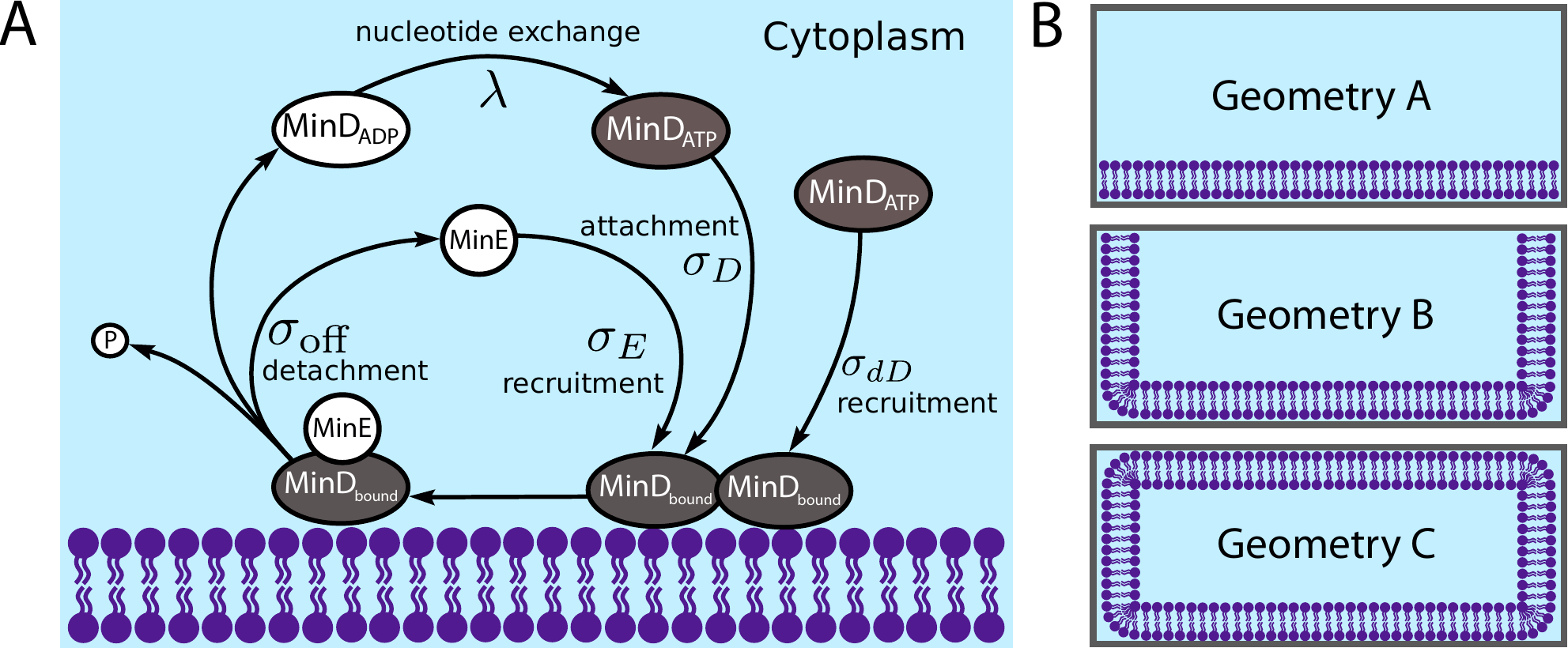}
	\caption{\textbf{A} Reaction cycle of the Min-system. The full cycle involves membrane attachment, 
		cooperative recruitment, diffusion along the membrane,
		detachment from the membrane, diffusion and a nucleotide exchange step in the cytosolic bulk. 
		\textbf{B} Cross sections of the three simulation geometries used in this paper. 
		In geometry A only the bottom is covered with a membrane. 
		Geometry B additionally covers the side walls and in geometry C the full compartment has covered boundaries.}
	\label{cycle}
\end{figure}

	On the molecular level, the oscillations emerge from the cycling of the ATPase MinD between 
	a freely diffusing state in the cytosolic bulk and a membrane-bound state, induced by its activator MinE under continuous consumption of chemical energy by ATP-hydrolysis, as shown schematically 
	in figure \ref{cycle}\hyperref[cycle]{A}. In its ATP-bound form MinD homodimerizes and can subsequently attach to the inner bacterial membrane as ATP-bound dimers using a membrane targeting sequence in form of a C-terminal amphipathic helix \cite{Lackner2003, Szeto2002, Szeto2003}. Despite the fact that the physicochemical details of MinD membrane binding are not yet fully understood, it has been demonstrated that MinD membrane binding is a cooperative process \cite{Lackner2003, Renner2012}. When being bound, MinD diffuses
	along the membrane. It also recruits MinE which in turn triggers the ATPase activity of MinD, breaking the complex apart and releasing all constituents back into the cytosol. MinD then freely diffuses in the bulk and can, after renewed loading of ATP and dimerization, rebind to the membrane at a new position. 
	This cycling of MinD between two states is the core mechanism for wave propagation of the Min-proteins. For a comprehensive overview on the underlying molecular processes, we refer to recent reviews on this topic \cite{Lutkenhaus2007, Lenz2011, Loose2011a, Vecchiarelli2012, Kretschmer2014,Kretschmer2016}.
	
	One of the most intriguing aspects of the Min-system is the impact of geometry on the spatiotemporal patterns. While initial experiments in wild-type cells showed characteristic \textit{pole-to-pole oscillations}, growing \textit{E. coli} cells, which roughly double in length before division, can also give rise to stable oscillations in both daughter cells even before full septum closure \cite{DiVentura2011}. In very long filamentous mutants the pole-to-pole pattern vanishes and several MinDE localization zones emerge in a stripe-like manner (\textit{striped oscillations}), with a characteristic distance of \SI{5}{\micro\metre}, strongly reminiscent of standing waves \cite{Raskin1999}. No stable oscillation patterns emerge in spherical cells where MinDE localization appears to be random without stable oscillation axes \cite{Corbin2002}.
	
	Strikingly, the Min-system can be reconstituted outside the cellular context using purified components on supported lipid bilayers \cite{Loose2008, Loose2011}.
	Using only fluorescently labeled MinD and MinE and ATP as energy source, traveling surface waves were observed in the form of turning spirals and traveling stripes on flat homogeneous substrates, where MinD proteins form a moving wave-front, that is consumed by MinE at the trailing edge, demonstrating that MinD and MinE alone are indeed sufficient to induce dynamic patterning \cite{Loose2008}. Interestingly, these assays work for different lipid species, demonstrating the robustness of the
	Min-oscillations with respect to the detailed values of the binding rates.
	
	Combining this reconstitution approach with membrane patterning, it was shown that the Min-system is capable of orienting its oscillation axis along the longest path in the patch and hence in principle capable of sensing the surrounding geometry
	\cite{Schweizer2012}. More recently the gap between the traveling \textit{in vitro} Min-waves and the standing Min-waves in live cells was closed, using microfabricated PDMS compartments mimicking the shape of \textit{E. coli} cells \cite{Zieske2013}. In these biomimetic compartments, which confine the reaction space in 3D, pole-to-pole oscillations were observed, reminiscent of the paradigmatic \textit{in vivo} oscillation mode. Later it was shown that the Min-oscillations are indeed sufficient to spatially direct FtsZ-polymerization to midcell, linking two key elements of bacterial cell division in a synthetic bottom-up approach \cite{Zieske2014}.
	
	In order to study the effect of geometry in the physiological context of the cell, one can place growing cells in microfabricated chambers of custom shape \cite{Mannik2012, Wu2015}. This \textit{cell sculpting} approach allowed the authors to systematically analyze the adaptation of the Min-oscillations to compartment geometry and demonstrated experimentally that different oscillation patterns can be stable for the same cell geometry. Using image processing, it was possible to measure the relative frequency of the different modes for a large range of interesting geometries \cite{Wu2015}. Figure \ref{cycle}\hyperref[cycle]{B}
	summarizes the different geometries that have been used
	before in experiments and that are considered here with computer simulations. While geometry A
	uses a flat membrane patch, similar to flat patterned substrates \cite{Schweizer2012}, geometry B corresponds to
	microfabricated chambers with an open upper side \cite{Zieske2013,Zieske2014}. Geometry C corresponds to the cell
	sculpting approach \cite{Mannik2012, Wu2015}.
	
	Like for other pattern forming systems, the theory of reaction-diffusion processes offers a suitable framework to address the Min-oscillations from a theoretical point of view \cite{Turing1952, Gierer1972, Cross1993, Kondo2010}. Many theoretical models have been proposed to unravel the physical principles behind this intriguing self-organizing protein system and to explain the origin of its rich spatiotemporal dynamics. While all of them rely on a reaction-diffusion mechanism 
	similar to the Turing model, they differ severely in their details. The first class of mathematical models used an effective one-dimensional PDE-approach and relied strongly on phenomenological non-linearities in the reaction terms \cite{Meinhardt2001, Howard2001, Kruse2002}. Although all of them successfully gave rise to pole-to-pole oscillations, they did not allow a clear interpretation of the underlying biomolecular processes and were not in agreement with all experimental observations, such as MinE-ring formation and the dependence of the oscillation frequency on biological parameters.
	
	The next advance in model building was the focus on the decisive role of MinD aggregation and the relevance 
	of MinD being present in two states (ADP- and ATP-bound) \cite{Kruse2002, Meacci2005}. 
	Very importantly, this highlighted the interplay between unhindered diffusion with a nucleotide exchange reaction 
	in the bulk as a delay element for MinD reattachment \cite{Huang2003}. 
	Subsequent models shared a common core framework but still differed strongly in the functional form 
	of the protein binding kinetics and the transport properties of membrane-bound molecules 
	\cite{Howard2005, Kruse2007}. 
	The main difference between the more recent models was the dimensionality, ranging from one dimension 
	\cite{Meinhardt2001, Howard2001, Kruse2002, Petrasek2015} to two \cite{Schweizer2012} and three 
	\cite{Huang2003, Kerr2006, Fange2006, Tostevin2006, Pavin2006, Wu2016}. 
	Moreover, the models can be classified as deterministic PDE-models \cite{Meinhardt2001, Howard2001, Kruse2002, 
		Huang2003, Meacci2005, Drew2005, Halatek2012, Bonny2013, Loose2008, Schweizer2012, Wu2015, Wu2016} 
	or using stochastic simulation frameworks \cite{Howard2003, Kerr2006, Fange2006, Pavin2006, Tostevin2006, 
		DiVentura2011, Bonny2013, Hoffmann2014}, and whether they neglected membrane diffusion 
	\cite{Meinhardt2001, Howard2001, Huang2003, Drew2005, Kerr2006} or not. 
	While some models contained higher than second order non-linearities in concentrations of the reaction terms 
	\cite{Loose2008}, it is the prevailing opinion to rely on at most second order non-linearities, 
	allowing for a clear interpretation in terms of bimolecular reactions. 
	Following the same line of thought, a strong effort was made to distill a minimal system that explains 
	the oscillation mechanism without the necessity of spatial templates or prelocalized determinants 
	\cite{Drew2005} and neglecting secondary processes like filament formation 
	\cite{Tostevin2006, Pavin2006, Drew2005}.
	
	The most influential minimal model for the Min-system has been suggested by Huang and coworkers \cite{Huang2003}. 
	It has been further simplified by discarding cooperative MinD recruitment by MinDE complexes on the membrane 
	\cite{Fange2006, Halatek2012}, allowing a clear view on the core mechanisms: 
	the cycling of MinD between bulk and membrane, cooperativity of MinD-recruitment and diffusion in bulk and along the membrane. Using the minimal model, it has been shown that the canalized transfer of proteins from one polar zone to the other underlies the 
	robustness of the Min-oscillations \cite{Fange2006, Halatek2012}. 
	Because the deterministic variants of the minimal model \cite{Huang2003, Halatek2012} do not allow us to address 
	the role of stochastic fluctuations, a stochastic and fully three-dimensional version has been introduced 
	to study the effect of stochastic fluctuations in patterned environments \cite{Hoffmann2014}. 
	For rectangular patterns of \SI{5}{\micro\metre} $\times$\SI{10}{\micro\metre}, it was found that the system 
	can be bistable, with transverse pole-to-pole oscillations along the minor and longitudinal striped 
	oscillations along the major axis, respectively. In this early work, it was observed that the stable phase
	emerged depending on the initial conditions and that sometimes switching occurred, but the statistics was not sufficiently good to observe switching in quantitative detail.
	Indeed such multistability has been observed experimentally in sculptured cells over a large range of cell
	shapes \cite{Wu2015} and the deterministic minimal model has been used to explain the relative frequency 
	of the different oscillations patterns for a given shape using a perturbation scheme \cite{Wu2016}. 
	However, as a deterministic model, this approach was not able to address the rate with which one pattern stochastically switches into another.
	
	Here we address this important subject by using particle-based Brownian dynamics computer simulations.
	Compared to earlier work along these lines 
	\cite{Hoffmann2014}, we have developed new methods to efficiently simulate and analyze the switching process. 
	We find excellent agreement with experimental data and measure for the first time the switching time of 
	multistable oscillation patterns. 
	We also use our model to confirm that it contains the minimal ingredients for the emergence of Min-oscillations.
	In addition, we use our stochastic model to investigate the three-dimensional concentration profiles in 
	different geometries and in particular the role of edges in membrane-covered compartments. We identify novel oscillation patterns
	in compartments with membrane-covered walls and find a surprisingly
	simple (linear) relation between the bound Min-protein densities
	and the volume-to-surface ratio, which might be relevant for
	geometry sensing by \textit{E. Coli} cells.
	
	\section{Methods}
	
\subsection{Reaction-diffusion model and parameter choice}
For the particle-based simulation, we use the reaction scheme of the minimal model for cooperative attachment \cite{Huang2003, Halatek2012, Hoffmann2014, Wu2016}. The model uses the following interactions between Min-proteins and the inner bacterial membrane (schematically shown in figure \ref{cycle}\hyperref[cycle]{A}). Freely diffusing cytoplasmic \un{MinD}{ATP} can bind to the membrane with a rate constant $\sigma_D$

\begin{subequations}
\begin{equation}
\text{MinD}_\text{ATP} \stackrel{\sigma_D}{\longrightarrow} \text{MinD}_\text{bound}.
\label{eq:reactionschemefirst}
\end{equation}

\un{MinD}{ATP} binds preferably to high density \un{MinD}{bound} regions (cooperative MinD binding)
\begin{equation}
\text{MinD}_\text{ATP} + \text{MinD}_\text{bound} \stackrel{\sigma_{dD}}{\longrightarrow} 2\  \text{MinD}_\text{bound}.
\end{equation}

Membrane-bound MinD also recruits cytoplasmic MinE to the membrane with rate $\sigma_E$, creating a \un{MinDE}{bound} complex
\begin{equation}
\text{MinD}_\text{bound} + \text{MinE} \stackrel{\sigma_E}{\longrightarrow} \text{MinDE}_\text{bound}.
\end{equation}
All membrane-bound proteins diffuse in the plane of the membrane,
but with a much smaller diffusion constant than in the bulk.

\begingroup
\begin{table}[b]
	\small
	\begin{ruledtabular}
	\begin{center}
		\item[]
		\caption{Parameters sets used for simulations of the Min-system. Set \textbf{A} is the main parameter set used here following \cite{Huang2003, Hoffmann2014}. Parameter set \textbf{B} is taken from \cite{Wu2015} and \textbf{C} from \cite{Halatek2012, Wu2016}.}
		\label{tbl:parameterset}
		\begin{tabularx}{\textwidth}{@{\extracolsep{\fill}}llllll}			
			Parameter & Set \textbf{A} &  Set \textbf{B} & Set \textbf{C} & Unit & Description (reaction type)\\
			\hline
			$D_D$ & \num{2.5} & \num{16} & \num{16} & \si{\micro\metre^2/ \second} & bulk diffusion coefficient of MinD \\
			$D_E$ & \num{2.5} & \num{10} & \num{10} & \si{\micro\metre^2/ \second} & bulk diffusion coefficient of MinE \\
			$D_{\text{bound}}$ & \num{0.01} & \num{0.013} & \num{0.013} & \si{\micro\metre^2/\second} & membrane diffusion coefficient \\ \hline
			$\lambda$ & \num{0.5} & \num{1} & \num{6}  & \si{\second^{-1}} &first order, unimolecular\\
			$\sigma_D$ & \num{0.025} & \num{0.075} & \num{0.1} & \si{\micro\metre/\second} & first order, membrane attachment\\
			$\sigma_{dD}$ & \num{0.0149} & \num{0.05}$^{a}$ & \num{0.1} & \si{\micro\metre^3/\second} & second order, bimolecular\\
			$\sigma_{E}$ & \num{0.093} & \num{0.25}$^{a}$ & \num{0.435} & \si{\micro\metre^3/\second} & second order, bimolecular\\
			$\sigma_{\text{off}}$ & \num{0.7} & \num{0.33} & \num{0.5} & \si{\second^{-1}} & first order, unimolecular \\ \hline
			$c_D$ & \num{0.797} & \num{0.85} & \num{1.0} & \si{\micro}M & total MinD concentration \\
			$c_E$ & \num{0.207} & \num{0.31} & \num{0.5} & \si{\micro}M & total MinE concentration
			\\
		\end{tabularx}
		\item[\small $^{a}$ In \cite{Wu2015} the unit for these bimolecular reactions is specified as \si{\micro\metre^2 \second^{-1}}.]
	\end{center}
\end{ruledtabular}
\end{table}
\endgroup

MinE attachment activates ATP hydrolysis of MinD. The hydrolysis of ATP to ADP breaks up the membrane-bound complex and releases \un{MinD}{ADP} and MinE back into the cytoplasmic bulk with rate constant $\sigma_\text{off}$
\begin{equation}
\text{MinDE}_\text{bound} \stackrel{\sigma_\text{off}}{\longrightarrow} \text{MinE} + \text{MinD}_\text{ADP}.
\end{equation}

Finally \un{MinD}{ADP} exchanges ADP by another ATP molecule (nucleotide exchange) with the rate $\lambda$
\begin{equation}
\text{MinD}_\text{ADP} \stackrel{\lambda}{\longrightarrow} \text{MinD}_\text{ATP}.
\label{eq:reactionschemelast}
\end{equation}
\end{subequations}
This completes the reaction cycle.
In table \ref{tbl:parameterset} we list our parameter values as set A. For comparison, we also list parameter values used in other studies (set B in \cite{Wu2015} 
and set C in \cite{Halatek2012, Wu2016}).

\subsection{Simulation algorithm}

We use custom-written code to simulate the stochastic dynamics of the Min-system with very good statistics.
For all simulations we use a fixed discrete time step of $\Delta t=\SI{E-4}{\second}$. During every time step each particle is first propagated in space. Thereafter every particle can react according to the previously introduced Min reaction scheme \ref{eq:reactionschemefirst} -- \ref{eq:reactionschemelast}. The movement of both free and membrane-bound particles is realized through Brownian dynamics. Individual molecules are treated as point-like particles without orientation. Therefore we can monitor the propagation separately for each Cartesian coordinate. During a simulation step of $\Delta t$ the displacements of the diffusing particles with diffusion constant $D$ are drawn from a Gaussian distribution with standard deviation $\sigma_x = \sqrt{2D \Delta t}$ \cite{Ottinger1995} such that
\begin{eqnarray}
&x(t+\Delta t)=x(t) + X_\text{G}, \\
&p_{X_\text{G}}(x)= \frac{1}{\sqrt{4 \pi D \Delta t}}
\exp\left(-\frac{x^2}{4 D \Delta t}\right), \label{eq3}
\end{eqnarray}
where $p_{X_G}$ is the probability distribution of $X_G$. The same update step is used for the $y$ and $z$ direction. Free particles in the bulk of the simulated volume undergo three-dimensional diffusion with reflective boundary conditions at the borders of the simulation volume. The membrane-bound particles perform a two-dimensional diffusion on the membrane with a much smaller diffusion constant $D_\text{bound}$ (compare table \ref{tbl:parameterset}). Membrane-bound particles are allowed to diffuse between different membrane areas that are in contact with each other. 

The different reactions in the Min reaction scheme \ref{eq:reactionschemefirst} -- \ref{eq:reactionschemelast} can be classified into three different types (more details on the corresponding implementations are given in the supplementary information).
The first type considered here are first order reactions without explicit spatial dependence. The conversion of \un{MinD}{ADP} to \un{MinD}{ATP} and the unbinding of the \un{MinDE}{bound} complex from the membrane are of this type. Such reactions are treated as a simple Poisson process. For a reaction rate $\kappa$, the probability to react during a time step $\Delta t$ is given by
\begin{equation}
p_{\kappa} = 1 - \exp\left(-\kappa \Delta t\right).
\end{equation}

The second type is also a first order reaction, but with confinement to a reactive area at a border of the simulated volume. The membrane attachment of \un{MinD}{ATP} proteins is a reaction of this type. For a given reaction rate $\sigma$, we implement these reactions by allowing particles that are closer to the membrane than $d=\SI{0.02}{\micro\meter}$ to attempt membrane attachment with a Poisson rate $\kappa=\sigma/d$. This results in a reaction probability of
\begin{equation}
p_\sigma=1 - \exp\left(-\frac{\sigma}{d} \Delta t\right).
\end{equation}

The last reaction type is a second order reaction between free and membrane-bound particles. The cooperative recruitment of cytosolic \un{MinD}{ATP} and MinE to membrane-bound MinD are reactions of this third type. In our simulation we adopt the algorithm implemented in the software package Smoldyn,
which has been used earlier to simulate the Min-system \cite{Hoffmann2014}. 
This algorithm is based on the Smoluchowski framework in which two particles react upon collision \cite{Smoluchowski1917}. However, the classical treatment by Smoluchowski only considers diffusion-limited reactions and therefore assumes instantaneous reactions upon collision. In order to take finite reaction rates into account, one imposes a radiation boundary condition \cite{Collins1949, Berg1978, Agmon1990}. From the diffusion constant $D$, the reaction rate $\sigma$ and the simulation time step $\Delta t$, a reaction radius $r_\sigma$ is calculated \cite{Andrews2004}. Whenever a freely diffusing particle comes within the distance of $r_\sigma$ to a membrane-bound particle, the free particle reacts.
For intermediate values of $\Delta t$ (such as the time step of \SI{E-4}{\second} that we use for the Min-system) the value of $r_\sigma(D, \sigma, \Delta t)$ is obtained numerically \cite{Andrews2004}. Those numerical values are taken from the Smoldyn software. For example, for parameter set A the reaction radius for the rate $\sigma_{dD}$ is $r_{\sigma_{dD}} = \SI{0.0091}{\micro\metre}$, and for $\sigma_{E}$ it is $r_{\sigma_{E}} = \SI{0.0179}{\micro\metre}$.

In our simulations we use rectangular reaction compartments. We considered three different membrane setups as illustrated in figure \ref{cycle}\hyperref[cycle]{B}. To mimic \textit{in vitro} experiments, where substrates or open compartments are functionalized with a membrane layer \cite{Schweizer2012,Zieske2013,Zieske2014}, we place the reactive membrane at the bottom (geometry A) or at the side walls and the bottom of the simulation compartment (geometry B). To simulate rectangular shaped \textit{E. coli} cells, inspired by the cell sculpting approach from \cite{Wu2015, Wu2016}, fully membrane-covered volumes are used (geometry C). We refer to the long side of the lateral extension as the major or the $x$-axis, and the smaller side as minor or $y$-axis, and accordingly consider the compartment height to extend in the $z$-direction, aligning the rectangular geometry perpendicular with the coordinate frame.

In our simulations we investigate a wide range of compartment dimensions. For a simulation box with a length of \SI{10}{\micro\metre}, width of \SI{5}{\micro\metre} and height of \SI{0.5}{\micro\metre}, we use \num{6003} \un{MinD}{ATP} particles, \num{6003} \un{MinD}{ADP} particles and \num{3124} MinE particles as initial condition \cite{Hoffmann2014}. These particle numbers amount to a total MinD concentration of $0.797\mu$M and a MinE concentration of $0.207\mu$M. For other simulation compartment sizes we scale the particle numbers linear with the volume, since in experiments \textit{E. coli} bacteria typically have a constant Min-protein concentration \cite{Wu2015}.

\subsection{Identification of oscillation modes}
In our simulations of the Min-system different oscillation patterns emerge along the major or minor axis of the 
simulation compartment. In order to analyze the frequency of different modes and the stability of the oscillations 
in the large amount of simulation data, an oscillation mode recognition algorithm is needed. 
Therefore we monitor the MinD protein densities at the poles of the different axes over time. 
To determine the axis along which the oscillation takes place, we compare the Fourier transformation of the 
normalized densities over time ($\rho_{t_i}$ where i denotes the discretized time resolution). 
If an oscillation takes place, there is a dominant peak in the Fourier spectrum and the overall maximal 
amplitude of the Fourier spectrum is significantly higher than the one from the non-oscillating axis. 
The same Fourier spectrum is also used to determine the oscillation period $T$.
To differentiate between pole-to-pole oscillations and striped oscillations of a given axis in the system, we extract the phase difference between the density oscillation at the poles of the cell.

When identifying switching events, one has to be more careful because stochastic fluctuations
might lead to temporal changes that might be mistaken to be mode switches. For this purpose,
we therefore smoothen the data. In detail, 
we calculate the convolution $C_i$ between the densities over time and a Gaussian time window $G_i$
\begin{equation}
C_i = \sum_j \rho_{t_j} G_{i-j}\, \quad \text{where} \quad
G_i = \frac{1}{\sqrt{2 \pi}\omega} \exp\left(-\frac{(i\tau)^2}{2 \omega^2}\right).
\end{equation}
Here $\tau$ is the time between successive density measurements and we set $\omega = \SI{100}{\second}$ as width of the time window. The current oscillation mode is now determined from the convoluted densities $C_i$
and assigned to the time $i\tau$. In this way, only switches are identified that persist for a sufficiently long time.

\begin{figure*}[t]
	\centering
	\includegraphics{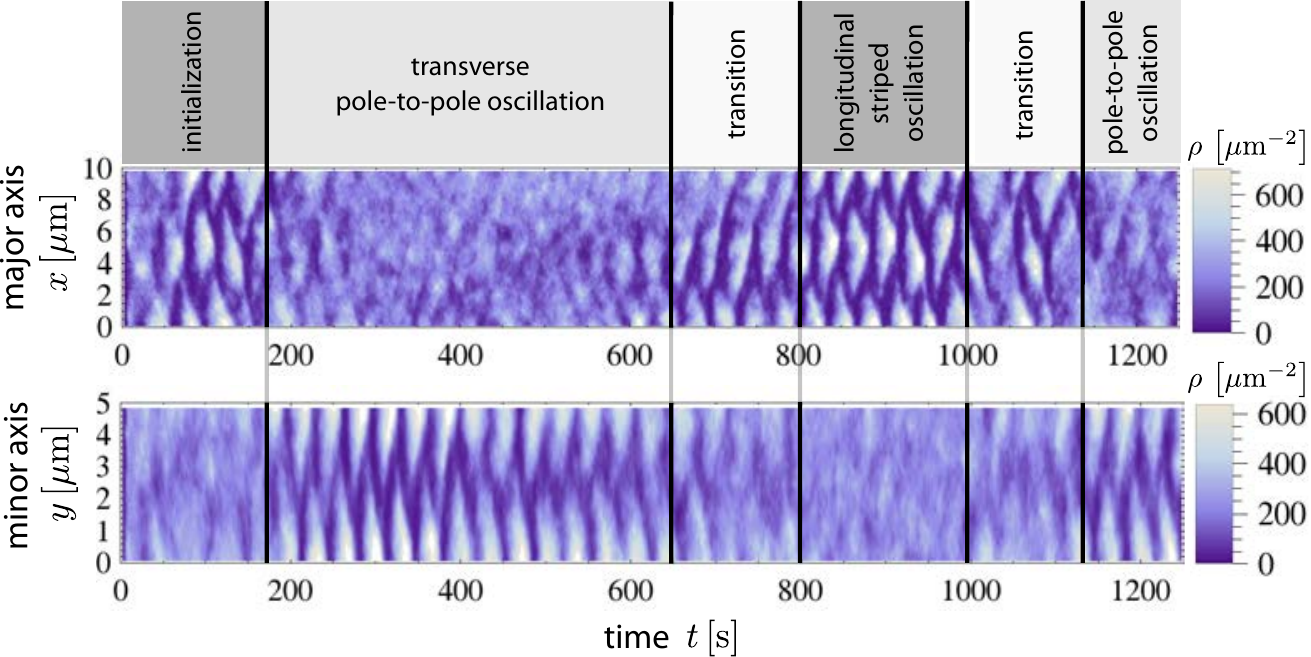}
	\caption{Density kymographs along both the major and minor axis of the system illustrating stochastic switching between a longitudinal striped oscillation mode (top) and a transverse pole-to-pole oscillation mode (bottom).
		Starting from an initially uniform particle distribution, first a longitudinal striped oscillation emerges along the major $x$-axis (top). After \SI{200}{\second} this oscillation stops and a transverse pole-to-pole oscillation along the minor $y$-axis begins (bottom). The oscillation mode switches again around \SI{700}{\second} and \SI{1100}{\second}. Here we use geometry A and the standard parameter set A. Dimensions are $\SI{10}{\micro\metre} \times \SI{5}{\micro\metre} \times \SI{0.5}{\micro\metre}$.
	}
	\label{switch}
\end{figure*}

\section{Results}

\subsection{Oscillation patterns in geometry A} \label{sec:geometryA}

First we investigated the oscillations that emerge in geometry A
with parameter set A using a rectangular simulation volume with dimensions $\SI{10}{\micro\metre} \times \SI{5}{\micro\metre} \times \SI{0.5}{\micro\metre}$ ($x,y,z$).
With this particular choice the width of the system approximately matches the typical length of wild-type  \textit{E. coli} cells and the length of the system corresponds to the length of a grown \textit{E. coli} cell which can roughly double in length before septum formation and division. 
As shown by the kymographs in figure \ref{switch} and in agreement with the results of Hoffmann \textit{et al.} \cite{Hoffmann2014}, in our simulations two different oscillation modes occur (compare also supplemental movie S1). 
Note from the color legend that dark and light colors correspond
to low and high concentrations, respectively, as used throughout
this work. 
In the first mode the Min-proteins oscillate along the minor $y$-axis from one pole to the other 
(pole-to-pole oscillation). 
In the second mode the proteins oscillate along the major $x$-axis between the poles and the middle of the 
compartment (striped oscillation). 
The system stochastically switches between the two modes, sometimes via a short
oscillation along the diagonal of the compartment.
The mode switching behavior of the Min-system in large volumes is in agreement with the experimental 
results of Wu \textit{et al.} \cite{Wu2015} and cannot be analyzed completely with conventional PDE-models of the Min-oscillations 
because they do not account for the noise in the system leading to the stochastic switch.

\begin{figure*}
	\centering
	\includegraphics[width=\textwidth]{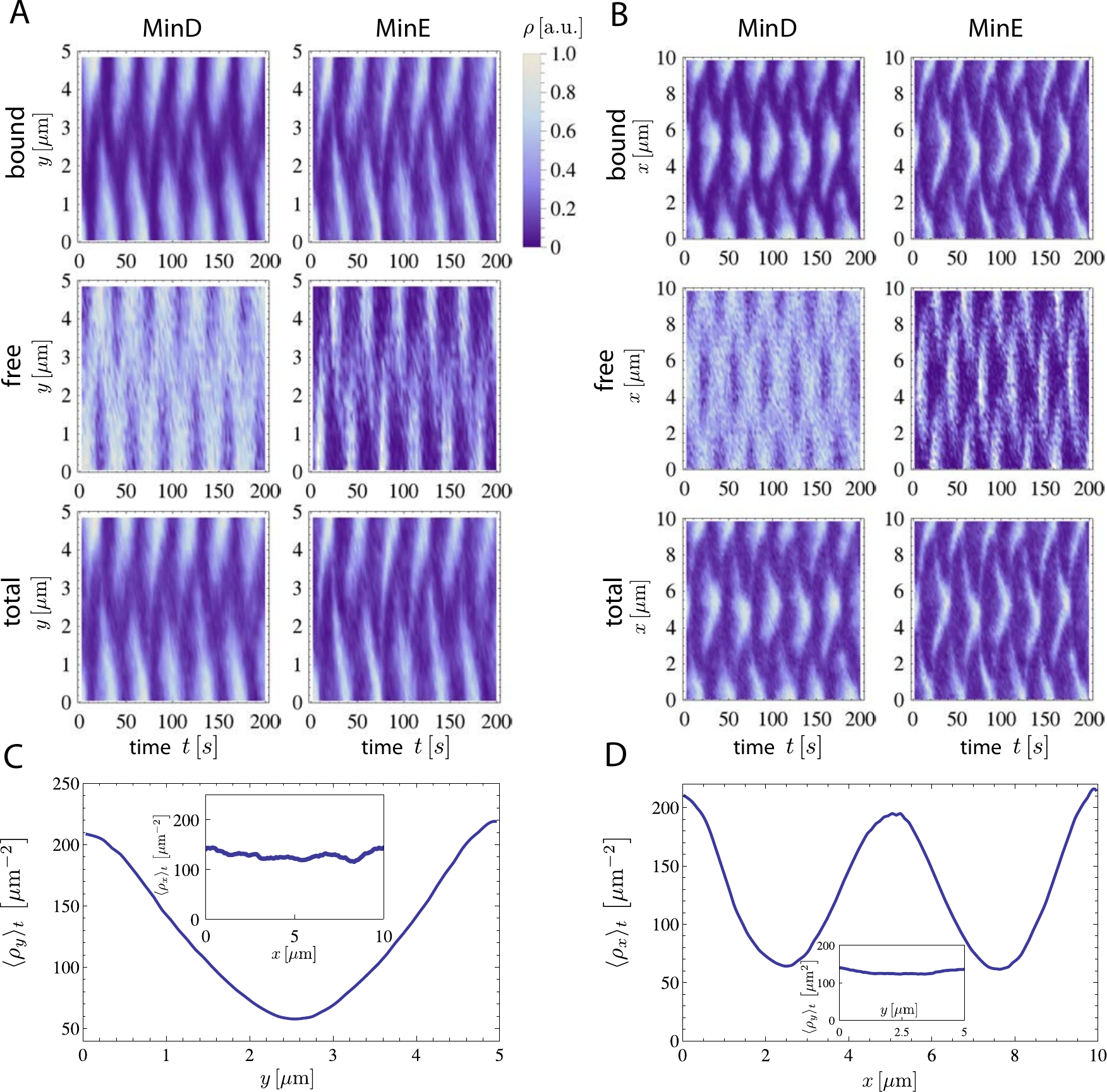}
	\caption{Detailed analysis of simulation results from figure \ref{switch}.
		\textbf{A} Kymographs of transverse pole-to-pole oscillations. Left column figures show MinD protein 
		particle densities and right column figures show MinE protein particle densities. The two figures on the top show 
		only particle densities of membrane-bound particles. The two figures in the middle row show particle 
		densities of free particles in the bulk of the simulation volume. The bottom two figures show the total particle 
		densities of both bound and free particles together. 
		\textbf{B} Kymographs of longitudinal striped oscillations. The arrangement is the same as in figure \ref{all}\hyperref[all]{A}.
		\textbf{C} Time-averaged density profile along the minor axis of bound MinD proteins during a pole-to-pole 
		oscillation (Inset: Time-averaged density profile along the major axis). 
		\textbf{D} Time-averaged density profile along the major axis of bound MinD proteins during a striped 
		oscillation (Inset: Time-averaged density profile along the minor axis). 
	}
	\label{all}
\end{figure*}

Detailed analyses of the pole-to-pole and the striped oscillations
from figure \ref{switch} are shown in figure \ref{all}\hyperref[all]{A} and \ref{all}\hyperref[all]{B}, 
respectively (compare also supplemental movies S2 and S3, respectively). 
First, we consider the pole-to-pole oscillations in figure \ref{all}\hyperref[all]{A}. In the kymographs of the bound MinD 
and MinE proteins (top row figures) we see clusters of bound proteins that detach from the membrane beginning in 
the middle and from there move towards one of the poles of the compartment in an alternating way. 
The shapes of the bound MinD protein density clusters in the kymographs have a triangular form, in contrast to the line-like structures of the
bound MinE proteins. Those density lines in the bound MinE kymograph show that the MinE proteins 
form a high density cluster in the middle of the cell which propagates to one of the compartment poles. 
This behavior is similar to the experimentally observed ringlike structures of MinE proteins in \textit{E. coli} 
bacteria
that travel from the middle to the poles of the cell, leading to the dissociation of MinD proteins from the membrane. 
The kymographs of the free particles (middle row in figure \ref{all}\hyperref[all]{A})
are averaged over all heights and have the inverse shape of the corresponding 
kymographs of the bound particles (top row in figure \ref{all}\hyperref[all]{A}). Where the density of bound particles is high, 
the density of free particles is low and vice versa. 
During the simulations the spatial density differences of both MinD and MinE are higher for membrane-bound particles than for free particles in the bulk. Therefore in the two bottom kymographs in figure \ref{all}\hyperref[all]{A}, which are showing total particle densities, the pattern of membrane-bound particles is dominant.

\begin{figure}[t]
	\centering
	\includegraphics{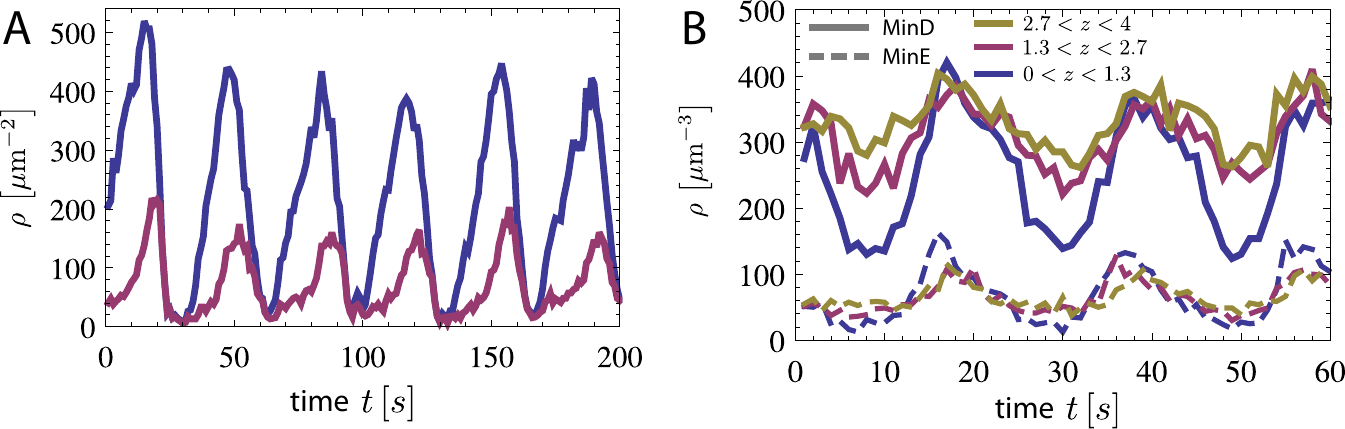}
	\caption{Differences between MinD and MinE. \textbf{A} Density change of MinD (blue) and MinE (red) in time during a pole-to-pole oscillation at position $y=\SI{4.9}{\micro\meter}$ using a compartment of $\SI{10}{\micro\metre} \times \SI{5}{\micro\metre} \times \SI{0.5}{\micro\metre}$ as in figures \ref{switch} and \ref{all}.
		\textbf{B} Density change of non-bound MinD and MinE in time at different heights above the membrane for a $\SI{6}{\micro\metre} \times \SI{3}{\micro\metre}$ bottom area and  $\SI{4}{\micro\metre}$ compartment height.
		\label{cut}
	}
\end{figure}

\begin{figure}[t]
	\centering
	\includegraphics{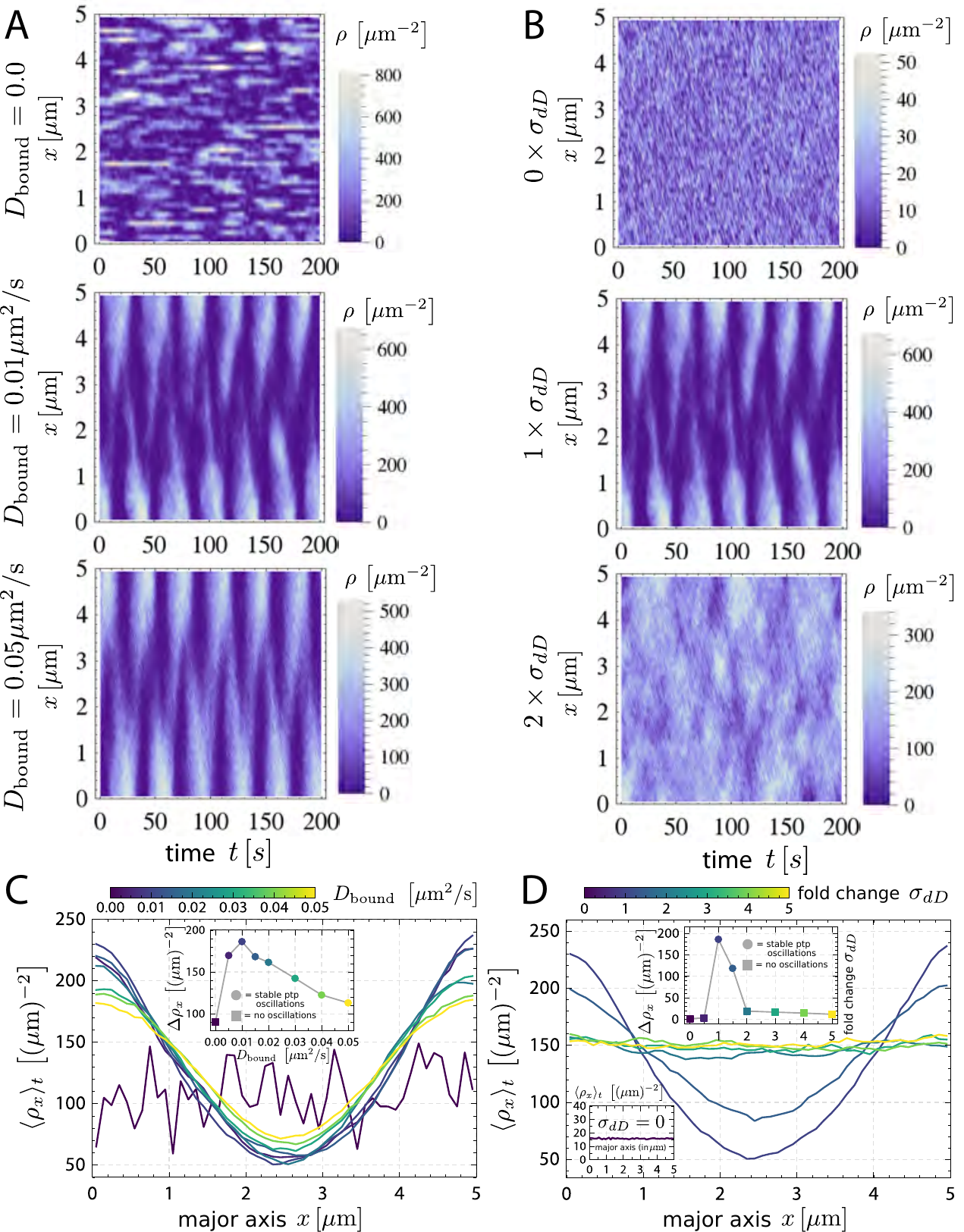}
	\caption{\textbf{A} Density kymographs along the major axis for $D_{\text{bound}}=0\ \si{\micro\metre^2/\second}$,
		$D_{\text{bound}}=0.01\ \si{\micro\metre^2/\second}$ and $D_{\text{bound}}=0.05\ \si{\micro\metre^2/\second}$, respectively.
		\textbf{B} Density kymographs along the major axis for $\sigma_{dD} = 0$, $\sigma_{dD} =  \SI{0.0149}{\micro\metre^3/\second}$ and  $\sigma_{dD} = \SI{0.0298}{\micro\metre^3/\second}$.
		\textbf{C} Time-averaged density profiles for different membrane diffusion-coefficients $D_{\textbf{bound}}$.
		\textbf{D} Time-averaged density profiles for different cooperative MinD membrane-recruitment rates $\sigma_{dD}$.
		\label{sensitivity}
	}
\end{figure}

The kymographs of the striped oscillation in figure \ref{all}\hyperref[all]{B} have the same structure as the ones of the 
pole-to-pole oscillations. However, the edges of the bound Min-protein clusters in the kymographs, that indicate 
the traveling Min-waves, are curved, in contrast to the straight lines that we see for the pole-to-pole oscillations as shown in figure \ref{all}\hyperref[all]{A}.
The time-averaged density profiles of MinD proteins for the pole-to-pole and striped oscillations are shown in figure \ref{all}\hyperref[all]{C} and \hyperref[all]{D}, respectively. As expected the density of the proteins is minimal between the oscillation nodes of the emerging standing wave patterns.

\begin{figure}[t]
	\centering
	\includegraphics{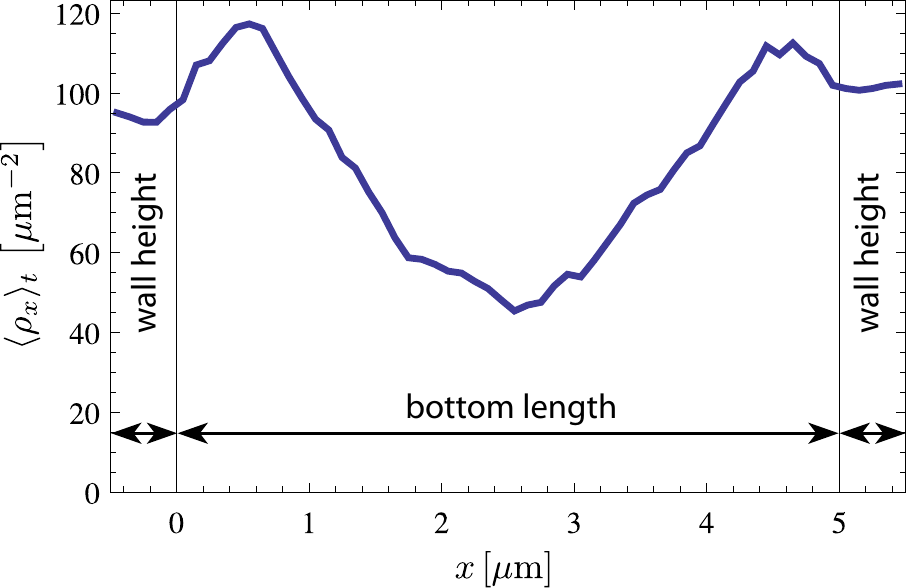}
	\caption{Average density profile of bound MinD proteins in a simulation where the reactive membrane is extended to the side walls of the compartment (geometry B). The figure shows the density profile on the bottom area of the compartment ($\SI{0}{\micro\meter} \le x \le \SI{5}{\micro\meter}$). Below $x=\SI{0}{\micro\meter}$ and above $x=\SI{5}{\micro\meter}$ the projected density profile along the side walls of the compartment is shown.}
	\label{densityprofileW}
\end{figure}

It is highly instructive to compare the time evolution of the MinD and 
MinE protein densities. In figure \ref{cut}\hyperref[cut]{A} we
plot the time evolution of the particle densities of the transverse pole-to-pole oscillation at a fixed position
$y=\SI{4.9}{\micro\metre}$, which is at the edge of the minor axis along which the oscillation takes place.
The shape of the transient density profiles is similar to experimentally observed density profiles of traveling Min-protein waves on flat membrane surfaces \cite{Loose2008, Loose2011}. The period of both oscillations modes was $T=\SI[separate-uncertainty=true]{33.8(1)}{\second}$, which is in agreement with the results of Huang \textit{et al.} \cite{Huang2003}.

To analyze the influence of the bulk volume on the oscillations, we have also monitored how the MinD and MinE densities changes at different heights above the membrane. For this we again use geometry A, but now with a bottom surface of only $\SI{6}{\micro\metre} \times \SI{3}{\micro\metre}$, which robustly produces longitudinal pole-to-pole oscillations along the major axis. For a compartment height of \SI{4}{\micro\metre} the volume was divided in three layers, and for each layer the mean density on one side of the major axis was plotted over time as shown in figure \ref{cut}\hyperref[cut]{B}. We see that the largest density changes take place in the layer directly above the membrane, however, the oscillations persist up to the top layer even for the highest compartment with $z=\SI{4}{\micro\metre}$. Strikingly, the
MinD-density is always much higher than the one of MinE. 
Furthermore we reduced the compartment height to \SI{0.2}{\micro\meter}. Interestingly, the pole-to-pole oscillations were still present. This implies that the bulk of the simulation volume has only a mild effect on the oscillations in geometry A, despite the fact that there are appreciable density 
variations in the bulk.

\subsection{Essential model elements}

We next checked that our model indeed includes the essential
elements for the emergence of Min-oscillations. Because the
MinD-switching between bulk and membrane is clearly indispensible, 
here we consider the relevance of diffusion on the membrane
and of cooperative recruitment. All simulations in this section are carried out using geometry A in a $\SI{5}{\micro\metre}\times \SI{2.5}{\micro\metre}\times\SI{0.5}{\micro\metre}$ compartment geometry.
We chose this geometry since we expect it to give rise to stable longitudinal pole-to-pole oscillations and hence can closely monitor any deviations from this default oscillation mode.
All other parameters were kept fixed and we used parameter set A for this test.

Figure \ref{sensitivity}\hyperref[sensitivity]{A} shows three kymographs of the system along the major axis. Switching off membrane diffusion entirely (first panel) leads to a loss of stable oscillation patterns. Instead small striped-like MinD-patches emerge erratically on the membrane.
The second panel in figure \ref{sensitivity}\hyperref[sensitivity]{A}  shows the default diffusion coefficient of $D_{\text{bound}}=0.01\ \si{\micro\metre^2/\second}$ where stable pole-to-pole oscillations emerge along the major axis. By further increasing the membrane-diffusion-coefficient the pole-to-pole oscillations still robustly emerge but increasingly smear out. This behavior is most clearly illustrated by looking at the time-averaged density profiles as shown in figure
\ref{sensitivity}\hyperref[sensitivity]{C}. The inset in figure \ref{sensitivity}\hyperref[sensitivity]{C} shows that without membrane diffusion, no oscillations emerge at all (square symbol), while a slow diffusivity as
used in parameter set A seems optimal.

In a similar fashion we analyzed the influence of the cooperative membrane recruitment of MinD. Figure \ref{sensitivity}\hyperref[sensitivity]{B} shows three kymographs for no cooperative recruitment ($\sigma_{dD}=0$), the default value from parameter set A and a two-fold increase, respectively. Without cooperativity in this process no oscillations emerge at all. The second panel shows again the default oscillation mode while already a two-fold increase also leads to unstable behavior without any patterns emerging. This sensitivity with respect to the cooperative recruitment rate $\sigma_{dD}$ is also summarized in figure \ref{sensitivity}\hyperref[sensitivity]{D}, which illustrates that only in a vary narrow range stable oscillations emerge. Although a complete
parameter scan is out of the question at the current stage for
reasons of computer time, we conclude that stable Min-oscillations
emerge only for certain parameter values and that parameter set A
performs very well in this respect.

\begin{figure*}
	\centering
	\includegraphics[width=0.95\textwidth]{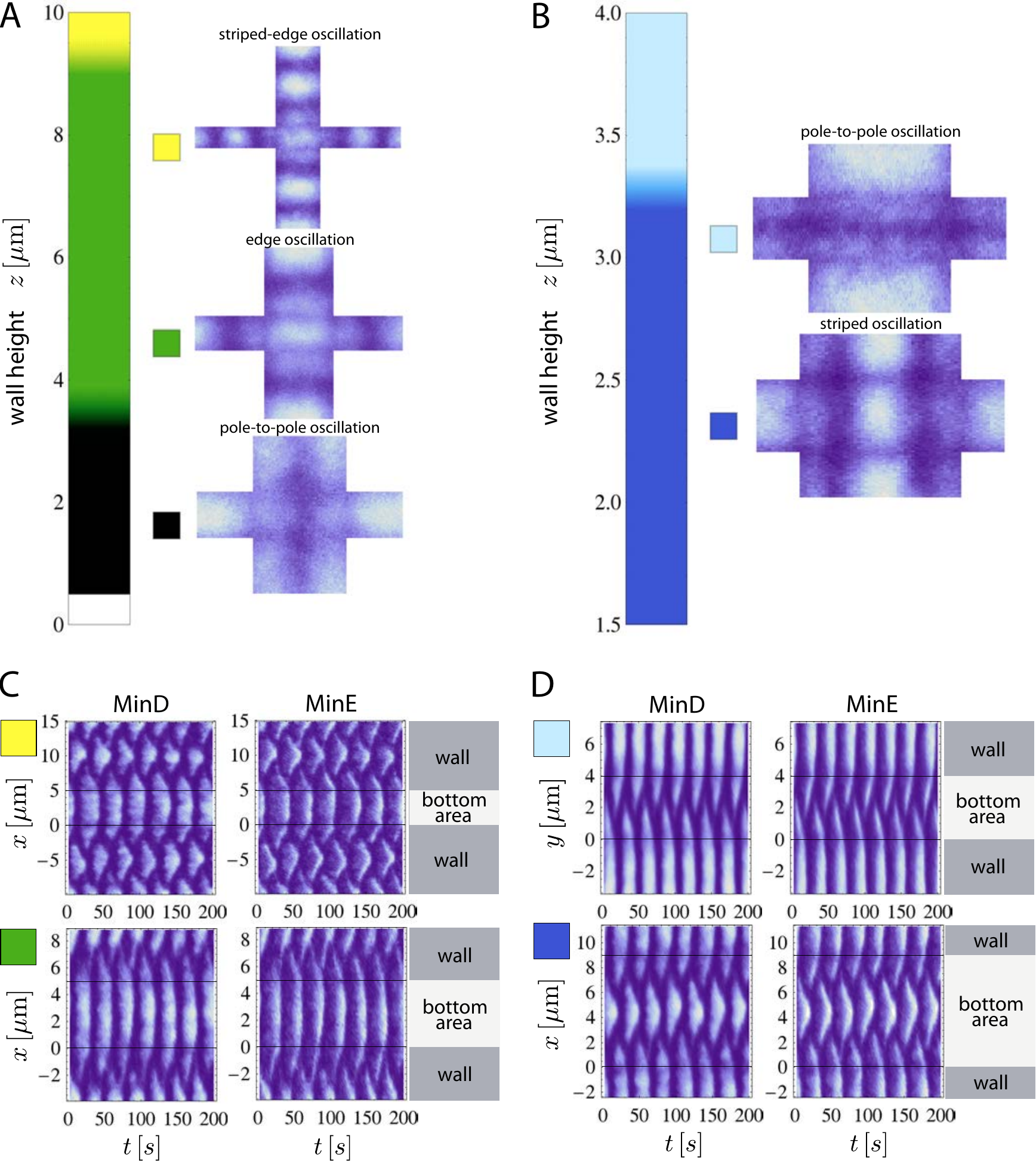}
	\caption{Different edge oscillation modes in geometry B in dependence of the wall height $z$ depicted as time-averaged density plots.
		\textbf{A} The bottom area has dimensions of $x=\SI{5}{\micro\metre}$ and $y=\SI{2.5}{\micro\metre}$. 
		Pole-to-pole oscillations are labeled in black, edge oscillations in green and the striped-edge oscillations in yellow.
		\textbf{B} The bottom area has dimensions of $x=\SI{9}{\micro\metre}$ and $y=\SI{4}{\micro\metre}$. 
		Striped oscillations are labeled in dark blue and pole-to-pole oscillations in light blue. In both A and B the density profiles show time-integrated surface-densities.
		\textbf{C-D} The corresponding kymographs show the densities of the bound MinD and MinE particles along the bottom area and the walls over time.}
	\label{phasediagram}
\end{figure*}

\begin{figure*}
	\centering
	\includegraphics[width=\textwidth]{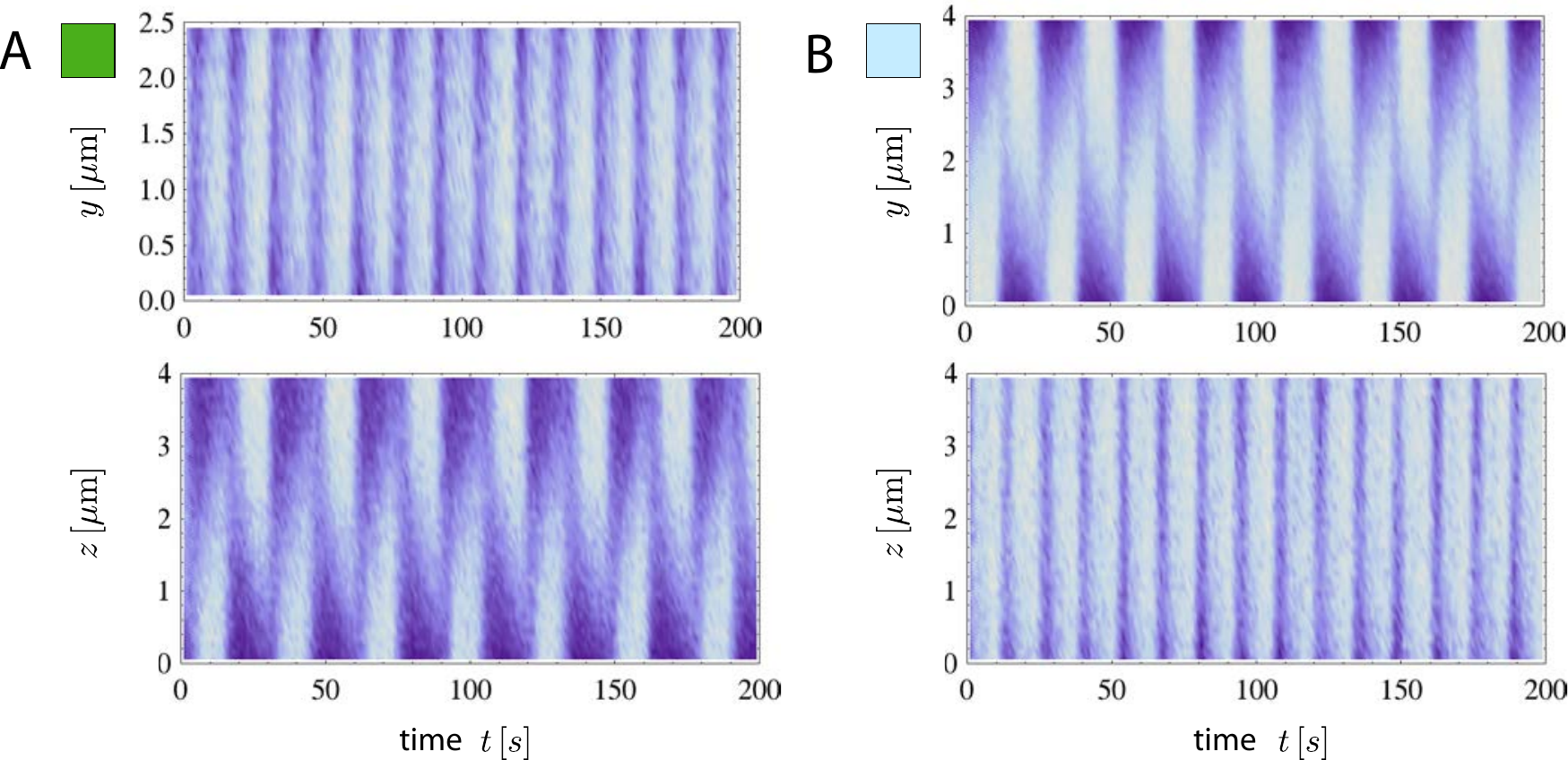}
	\caption{Density kymographs of free MinD particles in geometry B in the bulk of the compartments along different axes over time. \textbf{A} Edge oscillation (part of the green phase in figure \ref{phasediagram}\hyperref[phasediagram]{A}). \textbf{B} Pole-to-pole oscillation (part of the light blue phase in figure \ref{phasediagram}\hyperref[phasediagram]{B}).}
	\label{kymograph3d}
\end{figure*}

\subsection{Edge oscillations in geometry B}

As we have seen in the preceding section,
the simulation compartment height has little influence on the emerging oscillation modes
when the membrane only covers the bottom area of the simulation compartment (geometry A).
However, when we extend the membrane to cover bottom and side walls of the compartment (geometry B), we find that the oscillation modes change with increasing wall height. First, we consider a compartment that only exhibits longitudinal pole-to-pole oscillations along the major axis (wall height of \SI{0.5}{\micro\meter} and bottom area of $\SI{5}{\micro\metre}\times \SI{2.5}{\micro\metre}$).
In contrast to the flat membrane geometry A, here the MinD protein density decreases in the vicinity of the membrane edges between the side walls and the bottom area, as shown in figure \ref{densityprofileW}. This is due to the decreased volume per membrane area ratio in the vicinity of the membrane corners, leading to a decreased density of bound MinD proteins in these regions. This effect is clearly visible by comparing the time-averaged density profile in figure \ref{densityprofileW} with the one shown in \ref{all}\hyperref[all]{C}.

The full oscillation pattern is shown in \ref{phasediagram}\hyperref[phasediagram]{A} for the lowest height value (black label)
and resembles a pole-to-pole oscillation for geometry A, but now between the two opposing walls along the major axis.
Now we increase the wall height in increments of \SI{0.5}{\micro\metre}. The oscillation changes at a wall height of $\SI{3.5}{\micro\metre}$ to a new oscillation mode (green label, compare also supplemental movie S4). There the proteins start to oscillate between the middle of the bottom area and the upper edges of the walls.
This newly identified \textit{edge oscillation} with one polar zone at the bottom and one top polar
zone at each of the four side walls persists until the wall height reaches \SI{9.5}{\micro\metre}. 
Thereafter the oscillation mode changes to a striped edge oscillation along the walls (yellow label).

In figure \ref{phasediagram}\hyperref[phasediagram]{B} we show results for a simulation volume 
that gives rise to striped oscillations along the major axis (bottom area of the length of \SI{9}{\micro\metre} 
and width of \SI{4}{\micro\metre}). 
At a wall height of $z=\SI{0.5}{\micro\metre}$ the simulation gives rise to longitudinal striped oscillations 
along the major axis or transverse pole-to-pole oscillations along the minor axis. 
This is the same kind of bistability that we already observed in geometry A using a compartment of 
$\SI{10}{\micro\metre} \times \SI{5}{\micro\metre} \times \SI{0.5}{\micro\metre}$. 
After increasing the wall height to $z=\SI{1.5}{\micro\metre}$ the pole-to-pole oscillations along 
the minor axis disappear and only striped oscillations along the major axis are observed. 
At the wall height of $z=\SI{3.5}{\micro\metre}$ the oscillation mode changes to a large pole-to-pole 
oscillation along the minor axis and no edge or striped oscillations were observed.
Figures \ref{phasediagram}\hyperref[phasediagram]{C} and \ref{phasediagram}\hyperref[phasediagram]{D}
show the kymographs for the membrane-bound proteins corresponding to figures
\ref{phasediagram}\hyperref[phasediagram]{A} and \ref{phasediagram}\hyperref[phasediagram]{B}, respectively.
These results demonstrate that in geometry B, both the mode selection and the detailed 
shape of the polar regions can be controlled by compartment height.

For an edge oscillation the free MinD protein densities in the bulk of the compartment along one of the bottom area axis ($y$-axis) and along the walls ($z$-axis) are shown in figure \ref{kymograph3d}\hyperref[kymograph3d]{A}. We see that spatial oscillations in the bulk only take place along the $z$-axis. On the $y$-axis kymograph (figure \ref{kymograph3d}\hyperref[kymograph3d]{A} top kymograph) the change of the density only takes place along the temporal axis. This corresponds to the change of total numbers of free and bound MinD proteins during the oscillation and no spatial oscillation takes place along the $y$-axis. On the $z$-axis kymograph (figure \ref{kymograph3d}\hyperref[kymograph3d]{A} bottom kymograph) we observe a spatial pole-to-pole oscillation.
In figure \ref{kymograph3d}\hyperref[kymograph3d]{B} we show the densities of the free MinD proteins in the bulk 
of the compartment during the large pole-to-pole oscillation for $z=\SI{4}{\micro\metre}$. 
We see that the bulk proteins oscillate only along one axis (here the $y$-axis) and no spatial oscillations 
occur along the walls ($z$-axis).

From our study of geometry B, we conclude that in non-flat membrane geometries the oscillations of the Min-proteins do not only take place along the membrane, but are rather defined by the canalized transfer of the proteins through the bulk. This shows the importance of three-dimensional simulations for non-flat membrane geometries in order to determine the self-organized oscillation modes.

\subsection{Geometrical determinants of bound particle densities}

\begin{figure}[t!]
	\centering
	\includegraphics[width=0.95\textwidth]{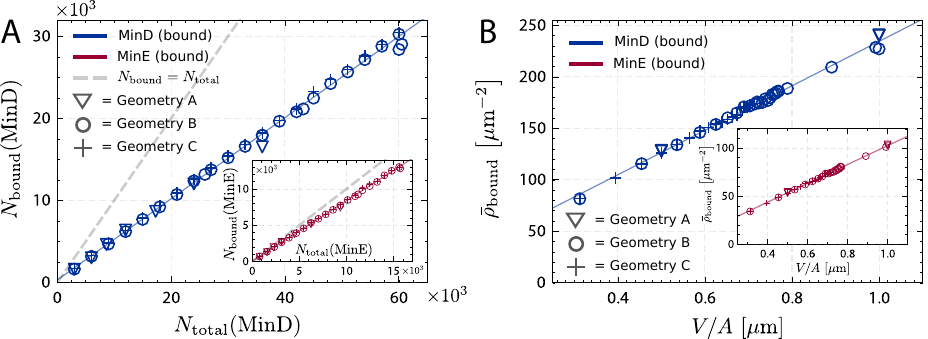}
	\caption{\textbf{A} Relation between the amount of total and membrane-bound MinD particles taken from many different independent simulations. The inset shows the same relation for bound MinE particles. 
		Note that MinE is more strongly depleted from the bulk
		than MinD (grey lines indicate complete depletion). 
		\textbf{B} Mean density of membrane-bound particles as a function of the $V/A$ ratios of the compartment geometry.}
	\label{NtoNb}
\end{figure}

We have also monitored the amount of membrane-bound 
MinD and MinE proteins and compared it to the total amount of proteins in the system. 
Interestingly, we observed that these two quantities have a linear dependence $N_\text{bound}\propto N_\text{total}$ 
as shown in figure \ref{NtoNb}\hyperref[NtoNb]{A}. 
Surprisingly, this relation seems to hold
for all geometries and dimensions, as indicated in figure \ref{NtoNb}\hyperref[NtoNb]{A} by the different symbols.
Since in our simulations we keep density constant, 
the total amount of proteins scales linearly with the compartment volume ($N_\text{total}\propto V$).
Overall, we conclude the following relation for the mean bound Min-protein density
\begin{equation}
\bar{\rho}_\text{bound}=\frac{N_\text{bound}}{A}\propto\frac{N_\text{total}}{A}\propto\frac{V}{A}
\end{equation}
where $A$ denotes the total reactive membrane area. 
Thus the mean bound Min-protein density increases linearly with the volume-to-area ratio, as verified by figure \ref{NtoNb}\hyperref[NtoNb]{B}.
Strikingly, we again observe a dramatic difference between
MinD and MinE. Although this scaling is the same for both,
MinE is much closer to being completely bound (grey lines), 
indicating that once a stable oscillation emerges, MinE is
almost completely depleted from the bulk.

\begin{figure}[t]
	\centering
	\includegraphics{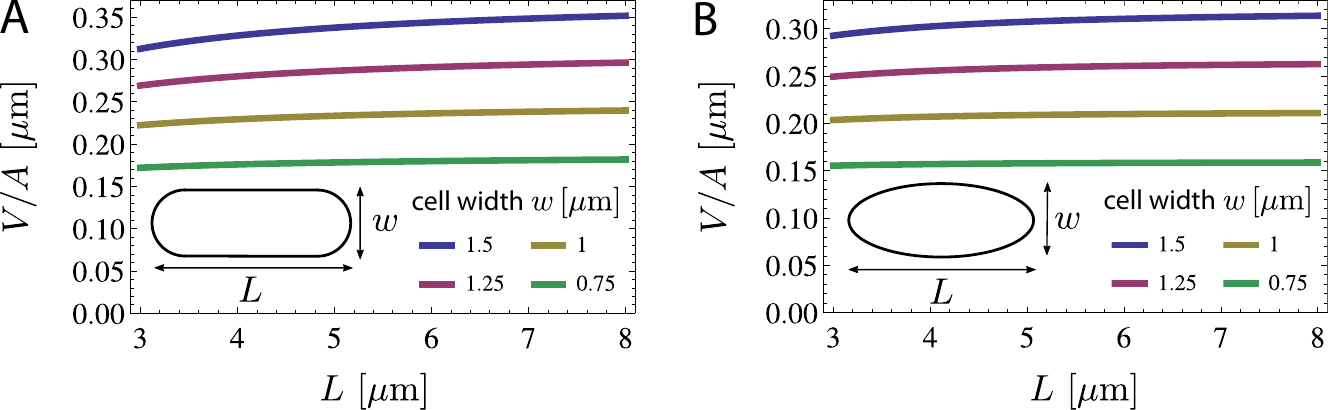}
	\caption{\textbf{A} Volume-to-area ratio $V/A$ of a spherocylinder as a function of the cell length $L$ for various cell widths $W$. 
		\textbf{B} A similar relation is observed for an ellipsoidal shape.}
	\label{VtoA}
\end{figure}

We next assessed the physiological relevance of these observations.
In order to investigate how the volume-to-area ratio $V/A$ changes during cell growth, we approximate the shape of an \textit{E.coli} bacterium
by a spherocylinder of length $L$ and width $W$ and evaluate $V/A$ analytically.
Since \textit{E. coli} bacteria mainly grow in length and stay constant in width \cite{Volkmer2011}, we show the evolution of $V/A$ as function of the cell length $L$ for various fixed cell widths in figure \ref{VtoA}\hyperref[VtoA]{A}.
One clearly sees that $V/A$ remains nearly constant as a function of the cell length $L$. Recalling that our previous observation stated that the mean membrane-bound Min-protein densities $\bar{\rho}_{\text{bound}}$ scale linearly with $V/A$, this would suggest that living bacteria grow in a fashion that keeps both $V/A$ and thus consequently $\bar{\rho}_{\text{bound}}$ constant, which might be advantageous for the stability and robustness of the Min-oscillations and related processes, such as formation
of the FtsZ-ring. A similar qualitative behavior is observed if we instead of a spherocylinder would assume an ellipsoidal shape as is shown in figure \ref{VtoA}\hyperref[VtoA]{B}.

\subsection{Oscillation mode switching}

\begin{figure*}
	\centering
	\includegraphics{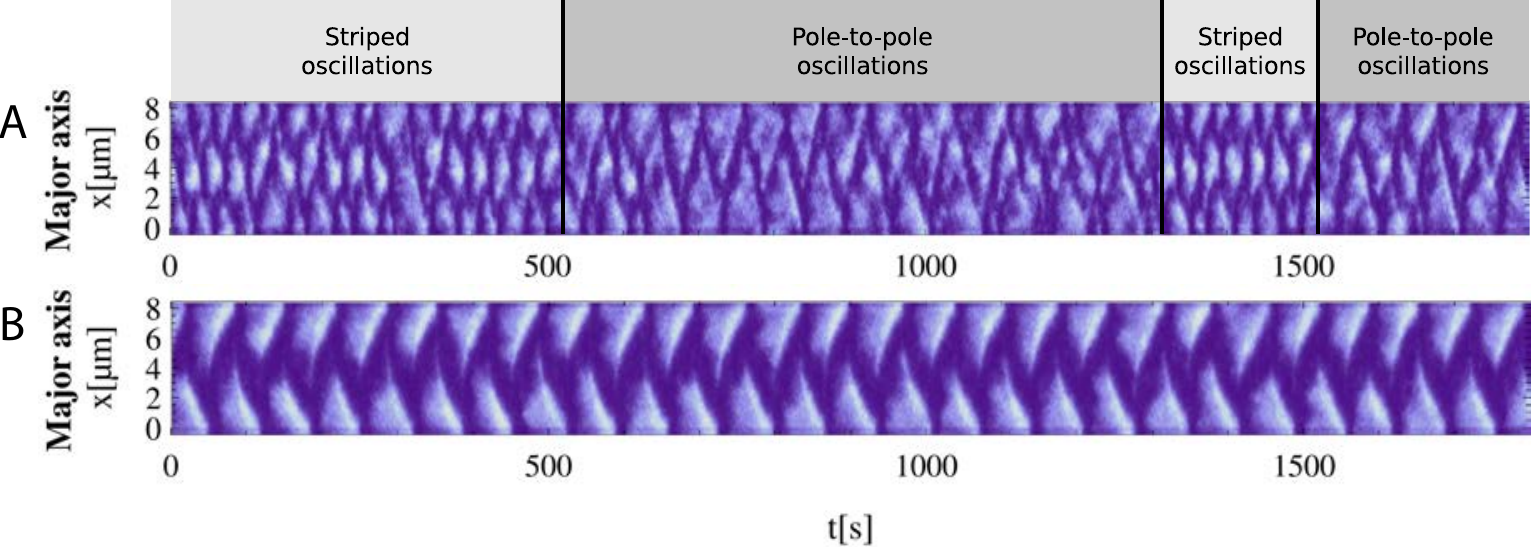}
	\caption{Stochastic oscillation mode switching in geometry B along the same axis. \textbf{A} shows the density kymograph of a simulation run using parameter set A where transitions between longitudinal pole-to-pole and longitudinal striped oscillations occur in a compartment of dimensions $\SI{8}{\micro\metre} \times \SI{2}{\micro\metre} \times \SI{0.5}{\micro\metre}$. \textbf{B} shows a density kymograph using the same compartment geometry but parameter set B instead.
	}
	\label{switch2}
\end{figure*}

\begin{figure}[t]
	\centering
	\includegraphics{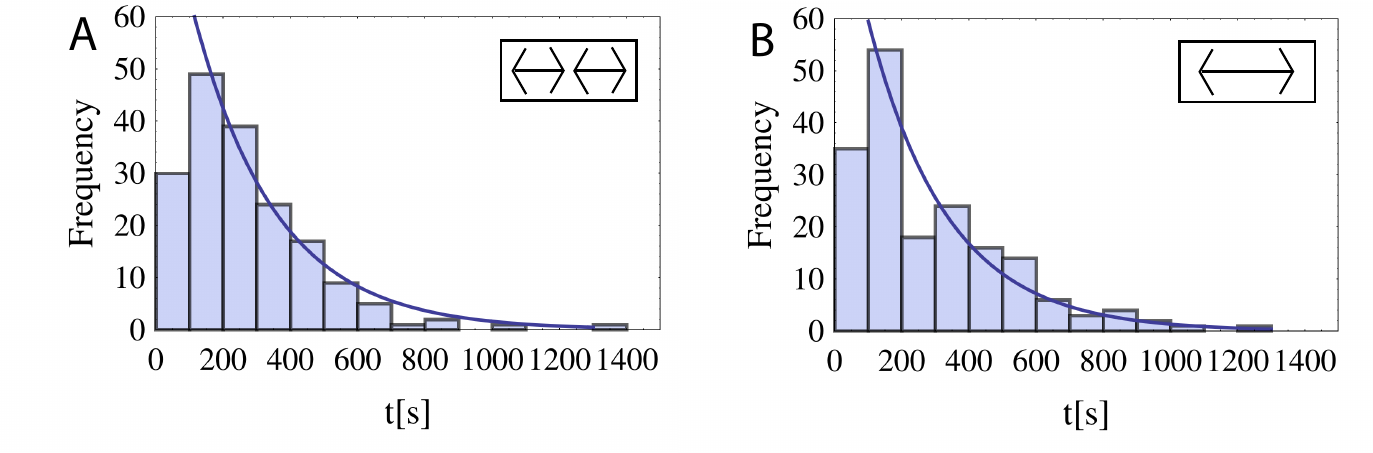}
	\caption{Histograms of oscillation mode lifetimes before a mode switch occurs. 
		\textbf{A} Striped oscillations.
		\textbf{B} Pole-to-pole oscillations.}
	\label{ratoi}
\end{figure}

Above we have seen that multistability is a recurrent phenomenon in the Min-system, both in geometries A and B.
We next turn to a systematic investigation of the stochastic switching between two different oscillation modes.
An oscillation mode transition of this type occurs frequently in a
$\SI{8}{\micro\metre} \times \SI{2}{\micro\metre} \times \SI{0.5}{\micro\metre}$
compartment using geometry B. In this geometry the Min-system gives rise to both pole-to-pole and striped 
oscillations along the same (major) axis (kymograph in figure \ref{switch2}\hyperref[switch2]{A}). 
We measure the lifetimes (oscillation duration before a mode switch occurs) of the two modes during a 
\SI{50000}{\second} long simulation trajectory. The resulting histograms of the lifetimes are shown 
in figure \ref{ratoi}. We can approximate the mode switching as a Poisson process by assuming that the switching 
probability $p(t)$ obeys $p(t) \propto \exp\left(-k t\right)$. Fitting an exponential to the switching times 
histograms we obtain the switching rates of these processes. Here we neglect the measurements of short lifetimes 
below $\tau_c  = \SI{100}{\second}$ since our oscillation mode detection algorithm can miss a mode transition 
if its lifetime is shorter. The switching rate for the pole-to-pole oscillation 
is $k_\text{p}=\SI{0.00422}{\second^{-1}}$ and for the striped oscillation we find 
$k_\text{s}=\SI{0.00406}{\second^{-1}}$, thus the two modes seem to be equally frequent
in this case.

We have also performed the same analysis with identical compartment geometry
for the two other parameter sets (set B and C) as presented in table \ref{tbl:parameterset}. 
Parameter set B gives rise to stable longitudinal pole-to-pole oscillations along the major axis 
(kymograph in figure \ref{switch2}\hyperref[switch2]{B}) in agreement with the results from \cite{Wu2015},
while parameter set C shows a qualitatively similar switching behavior 
as parameter set A (data not shown) with frequent mode transitions.

\begin{figure}[t]
	\centering
	\includegraphics[scale=1.35]{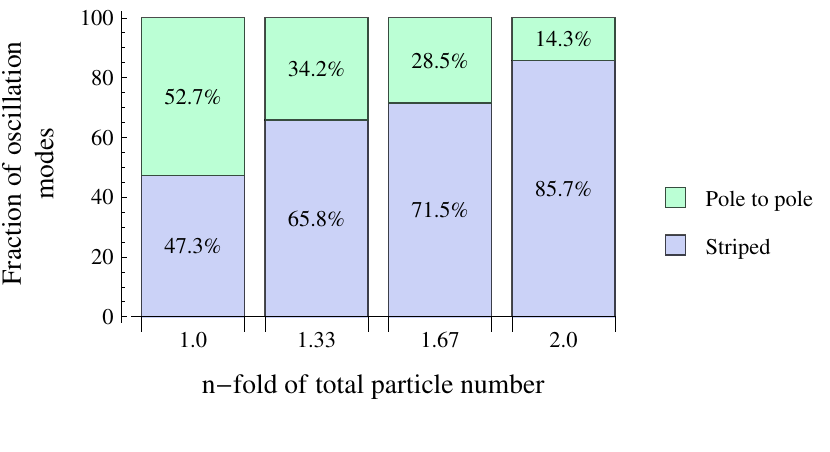}
	\caption{Fractions of pole-to-pole and striped oscillation modes in dependence of the amount of particles present in the simulation volume (geometry B, parameter set A).}
	\label{switchData}
\end{figure}

To analyze the effect of the Min-protein concentration on the oscillation mode switching, we have performed the same simulation with increased Min-protein particle densities using  again parameter set A. The resulting fractions of the two oscillation modes during the \SI{50000}{\second} simulation trajectory are shown in figure \ref{switchData}, and their corresponding switching rates $k_p$ and $k_s$ are given in table \ref{tbl:rates}.
We note that the fraction of striped oscillations increases monotonously with increasing particle density. In agreement with this, the 
rate $k_s$ decreases with increasing particle density.
In contrast, the rate $k_p$ does not show a systematic change
and seems to fluctuate strongly. The shift to the striped
oscillations suggests that the Min-oscillations can be used
not only to sense geometry, but also to sense concentrations. 

\begingroup
\begin{table}[b]
	\small
	\begin{ruledtabular}
	\begin{center}
		\item[]
		\caption{Switching rates for different protein concentrations (geometry B, parameter set A).}
		\label{tbl:rates}
		\begin{tabularx}{0.55\textwidth}{@{\extracolsep{\fill}}cll}
			$n$-fold particle number & $k_p \, \left[\si{\second^{-1}}\right]$ &  $k_s  \, \left[\si{\second^{-1}}\right]$  \\ \hline
			$1$ & \num{0.0042} & \num{0.0041} \\
			$1.33$ & \num{0.0065} & \num{0.0024} \\
			$1.67$ &  \num{0.0055} & \num{0.0016} \\
			$2$ & \num{0.0066} & \num{0.0008}
		\end{tabularx}
	\end{center}
	\end{ruledtabular}
\end{table}
\endgroup

\subsection{Phase diagrams for geometry C}

Wu \textit{et al}. have experimentally measured the fractions of different oscillation modes of Min-proteins in 
rectangular cell geometries of \textit{E. coli} bacteria of various sizes with constant height \cite{Wu2015}. 
In order to address these observations, we now turn to geometry C,
which is a rectangular and fully membrane-covered compartment as sketched in figure \ref{cycle}\hyperref[cycle]{B}. 
Throughout this section we keep the compartment height fixed at $z =\SI{1}{\micro\metre}$, 
while we vary the length (major axis) between \num{2} and $\SI{10}{\micro\metre}$ and the width (minor axis) 
of the compartment between \num{1} and $\SI{5}{\micro\metre}$ in steps of $\SI{1}{\micro\metre}$, respectively. 
To determine the oscillation mode fractions we calculate an ensemble average of the oscillation modes after \SI{500}{\second} simulation time. 
For each compartment size a sample of 20 independent simulations is used.

\begin{figure}[t!]
	\centering
	\includegraphics[width=0.95\textwidth]{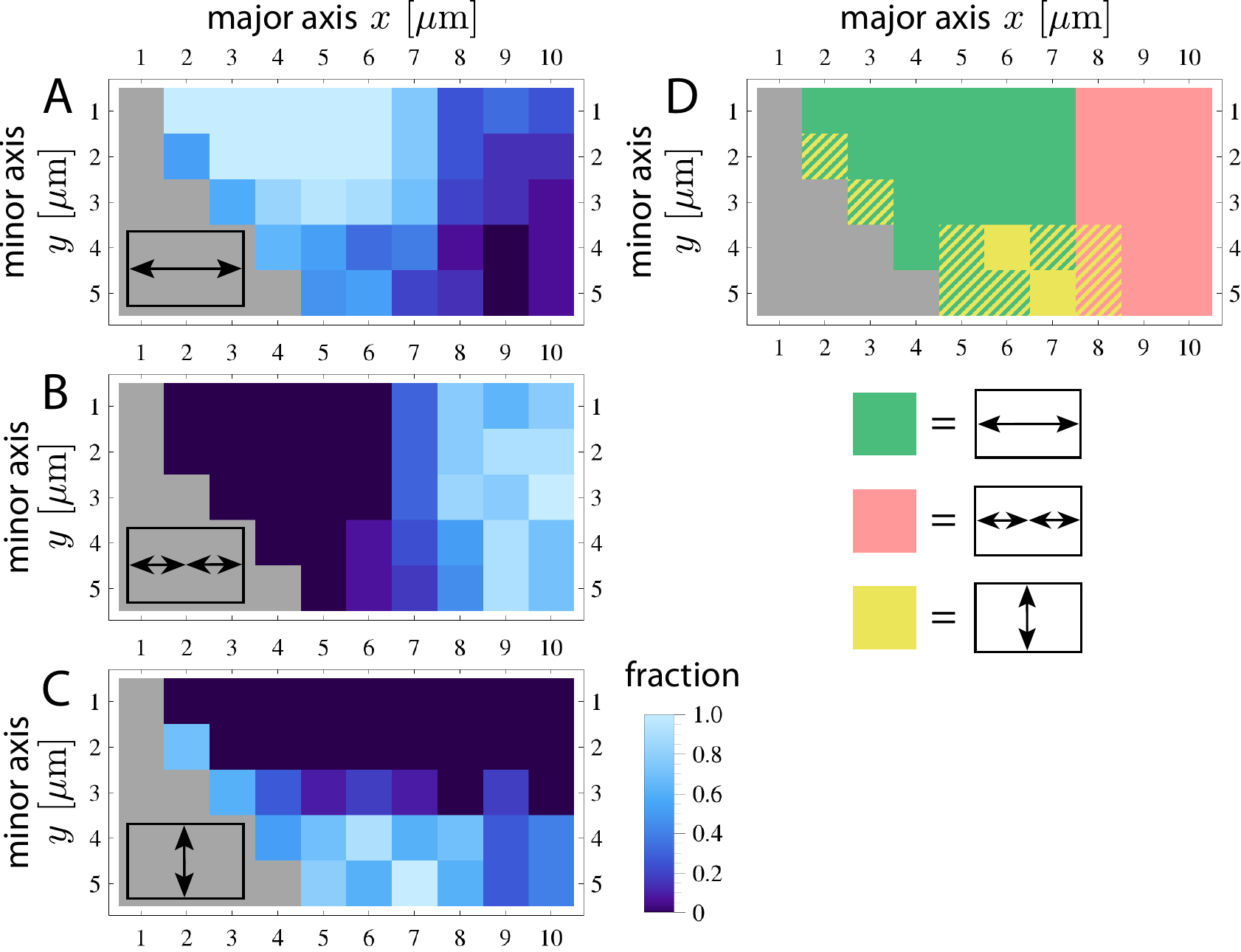}
	\caption{Relative importance of the three main oscillation modes in
		geometry C of varying dimensions. \textbf{A} shows the fractions of longitudinal pole-to-pole oscillations along the $x$-axis, \textbf{B} the fractions of longitudinal striped oscillations along the $x$-axis, and \textbf{C} transverse pole-to-pole oscillations along the $y$-axis.
		\textbf{D} Phase diagram of dominant oscillation modes. When two modes emerged both with a frequency > $40\%$, we considered both modes as dominant, as indicated by the striped regions in the diagram.}
	\label{fractions}
\end{figure}

In the following analysis we only focus on the three main oscillation modes, 
which here are longitudinal pole-to-pole and striped oscillation (along the major axis) and transverse 
pole-to-pole oscillation (along the minor axis). 
Other oscillation modes as they have been reported in \cite{Wu2015} were also observed using our framework, however, 
due to their low probability, a much higher sample size would be necessary to analyze their occurrence frequency 
with sufficient statistics. The results for the three oscillation modes are shown in figure \ref{fractions}\hyperref[fractions]{A-C}.
In general, each of the three oscillation modes dominates in one region of the phase diagram,
but the transitions are fuzzy and therefore bistabilities occur.
For compartments with length below $\SI{7}{\micro\metre}$ and width below $\SI{4}{\micro\metre}$, only pole-to-pole oscillations are observed. Most of those oscillations occur along the major axis of the system.
The transverse pole-to-pole oscillations along the $y$-axis emerge most frequently in compartments with quadratic bottom area.
Increasing the width further increases the fractions of transverse pole-to-pole oscillations at the expense of longitudinal pole-to-pole oscillations. 
Increasing the length for a fixed width shows a sharp transition from longitudinal pole-to-pole to longitudinal striped patterns at around \SI{6}{\micro\metre}.
For compartments with length larger than $\SI{7}{\micro\metre}$, the longitudinal pole-to-pole oscillations vanish almost entirely.
In the region of both large long sides (around \num{7} to \SI{10}{\micro\metre} in length) and large short sides (around \num{4} to \SI{5}{\micro\metre} in width), the oscillation mode fractions are rather equally shared between longitudinal striped and transverse pole-to-pole oscillations, which is also in line with the bistability that we observed between these two patterns in the $\SI{10}{\micro\metre} \times 
\SI{5}{\micro\metre} \times \SI{0.5}{\micro\metre}$ compartment of geometry A as shown in section \ref{sec:geometryA}. 
Figure \ref{fractions}\hyperref[fractions]{D} summarizes
these findings in a phase diagram that considers only the
dominating mode (expect in the regions of clear bistability).
Overall we find the determined oscillation mode fractions based on parameter set A and as shown in figure \ref{fractions} to be in excellent agreement with the experimental results as reported by Wu \textit{et al.} \cite{Wu2015}.

\section{Conclusion}

Using a stochastic particle-based simulation framework, we have investigated the stochastic switching between multistable
oscillation modes of the Min-system in different three-dimensional compartment geometries.
Although it is well known that geometrical constraints have a strong impact on the dynamic oscillations 
of the Min-proteins, multistability and mode switching have only recently been investigated in
more detail \cite{Hoffmann2014, Wu2016}.
Our stochastic framework provides a suitable approach to address the question of oscillation mode 
selection and stability, since it naturally incorporates fluctuations due to finite copy numbers in the system.
This allowed us to address the influence of the three-dimensional shape of the compartment and the boundary conditions,
with a close match to existing experimental assays (flat supported bilayers \cite{Loose2008,Schweizer2012}, functionalized compartments \cite{Zieske2013,Zieske2014}, and cell sculpting \cite{Wu2015,Wu2016}). 
For example, we addressed the role of compartment height and demonstrated the emergence
of new oscillation modes with increasing height, underlining the importance of an explicit three-dimensional
representation of the system. We also showed
that diffusion long the membrane and cooperative
recruitment are essential elements for the emergence of Min-oscillations.
In general, particle-based
stochastic computer simulations are a great tool for explorative
research and in the future could be used to explore more details of
the different scenarios that have been suggested for the molecular mechanisms shaping the Min-oscillations \cite{Bonny2013,Petrasek2015}.

Our simulations demonstrated several features that might be related to the physiological function of the
Min-system in \textit{E. coli}. First we observe that there is a dominating length scale of
$5\mu\text{m}$, which happens to be the natural length of an interphase \textit{E. coli} cell. Second
we found a linear relation between the density of membrane-bound Min-proteins and
the volume-to-surface ratio, which tends to be constant during growth of \textit{E. coli}.
Third we observed that the relative frequency of competing oscillation modes depends on
concentration, suggesting that the Min-oscillations can be used not only to sense geometry,
but also concentration. Fourth, we found that 
stable oscillations strongly deplete MinE from the bulk.
For the future, it would be interesting to study possible feedback
between protein production and Min-oscillations.

For our simulations, we mainly used the established parameter set A from table \ref{tbl:parameterset}
and achieved excellent agreement with experimental results regarding the 
relative frequency of the three main oscillation modes in different cell geometries \cite{Wu2015}.
Again the critical length scale around \SI{5}{\micro\metre} plays an important
role in transitions between different regimes, which then are the regions of high bistability.
However, we also note that different parameter choices lead to different outcomes
and that it would be interesting to perform an exhaustive exploration of parameter space
to better understand how robustness and multistability depends on kinetic rates,
diffusion constants and concentrations. In the future, such
a complete scan might become possible by using GPU-code rather
than the CPU-code developed here.

Most importantly, however, our stochastic approach allowed us to measure for the first
time the switching rates between different competing oscillation patterns of the Min-system.
This was done for parameter set A from table \ref{tbl:parameterset}. Interestingly, parameter
set C gave similar results in this respect, while parameter set B results in very stable oscillations without
switching, in agreement with the experimental observations for the cell sculpting experiments
\cite{Wu2015}. In the future, it would be interesting to investigate more systematically
how the effective barriers between two competing oscillation patterns depend on model
parameters and compartment geometry. In general, the Min-system is an excellent model
system to study not only geometry sensing, but also the role of spatiotemporal fluctuations
in molecular systems.

\begin{acknowledgments}
USS is a member of the Interdisciplinary Center for Scientific Computing (IWR) and the CellNetworks cluster 
of excellence at Heidelberg University. NDS acknowledges support by the BMBF program ImmunoQuant and by the 
German Academic Scholarship Foundation.	
\end{acknowledgments}

\appendix

\section{Particle interactions in the simulation algorithm}
\label{PInt}
The Min reaction scheme as introduced in equations (1a -- 1e) of the main text consists of five molecular reactions, which can be classified into three different types of reactions.
\begin{itemize}
	\item[($i$)] First-order unimolecular reactions of the type 
	$A \xrightarrow[]{\,\, k_1\,\,} B$. These kind of reactions have no dependence on the spatial coordinates of the system. In the Min reaction scheme the nucleotide exchange reaction ($\lambda$) in the bulk and the membrane detachment ($\sigma_{\text{off}}$) are of this type.
	\item[$(ii)$] Simple membrane attachment reactions are also first-order unimolecular reactions of the above type $A \xrightarrow[]{\,\, k_{1,m}\,\,} B$, but with the additional constraint that a particle must be in close proximity to a membrane to be able to attach to it.
	The binding of MinD in its ATP-bound state to the membrane ($\sigma_D$) is a reaction of this type.
	\item[$(iii)$] Bimolecular membrane attachment reactions. This reaction type is of second-order
	\begin{align}
	A + B\xrightarrow[]{\,\, k_{2,m}\,\,} \text{product(s)} 
	\end{align}
	and describes the bimolecular association of a particle in the bulk with an already membrane-bound particle. The cooperative MinD recruitment ($\sigma_{dD}$) and the MinE recruitment ($\sigma_E$) by membrane-bound MinD are reactions of this type. 
\end{itemize}
To create a consistent particle-based simulation algorithm for the above introduced types of reactions, we compare our reactive Brownian dynamics algorithm with the corresponding mean-field partial differential equations, as they are for example studied in \cite{Huang2003}. For this we consider each of the three reaction types individually using only a simplified minimal setting.

\subsection*{\textbf{$\bm{(i)}$} First-order unimolecular reactions}
For the first reaction type the corresponding differential equation is a simple Poisson process
\begin{align}
\partial_t\rho = - k_{1} \rho,
\end{align}
where $\rho$ denotes the free particle density and $k_{1}$ is the corresponding first-order reaction rate constant. Introducing a concrete system size $V$ we can equally convert particle densities into particle numbers $N$ and have equivalently $\partial_t N = -k_1 N$ with the standard exponential solution
\begin{align}
\rho(t) = \rho_0 e^{-k_1 t} \, , \qquad N(t) = N_0 e^{-k_1t}.
\label{eq7}
\end{align}
In a particle-based simulation framework with a total of $N_0$ particles of a generic species $A$ we are interested in the reaction probability $p_{\text{react,1}}(\Delta t)$, that a particle of type $A$ undergoes such a first-order reaction within a fixed, finite time step of length $\Delta t$. After a time $ t = \Delta t$ the solution of the linear degradation equation \eqref{eq7} tells us that $n_r$ particles will have reacted
\begin{align}
n_r = N_0 -  N(\Delta t) = N_0\bigl(1 - e^{-k_1\Delta t}\bigl).
\end{align}
This means that a single particle has a probability of
\begin{align}
p_{\text{react,1}}(\Delta t) = 1 - \exp(-k_1 \Delta t),
\end{align}
to react during a time step $\Delta t$.

\subsection*{\textbf{$\bm{(ii)}$} First-order membrane attachment}
For this second reaction type we consider a setting with a membrane at the bottom surface at $z=0$ of a rectangular simulation volume $V$ to which a generic species $A$ can attach.
In this scenario the lateral displacements in both the $x$- and $y$- direction can be integrated out in a straightforward way allowing us to reduce the system to an effective one-dimensional system where $z$ is the only spatial dependency that remains. In the resulting one-dimensional domain we have a
membrane at $z=0$ and impose a reflecting no-flux boundary condition at $z=L_z$.
The corresponding differential equation for the particle density $\rho(z,t)$ now reads
\begin{align}
\partial_t\rho(z,t) = D\partial_{zz}\rho(z,t) - \sigma \delta(z)\rho(z,t) \quad \text{with} \quad 0 \le z \le L_z
\label{eq8}
\end{align}
subject to $\left.\partial_z p(z,t)\right|_{z = L_z} = 0$.
\begin{figure}[h!]
	\centering
	\includegraphics[scale=1.2]{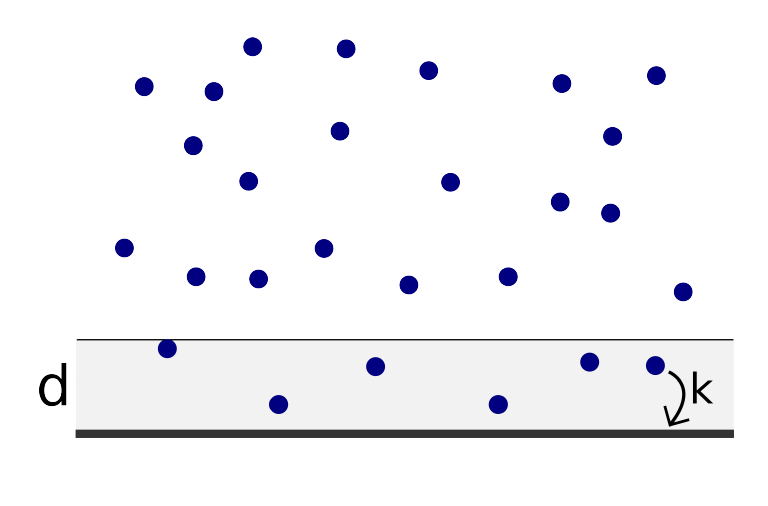}
	\caption{The model used for first-order membrane attachment reactions (type $(ii)$) in the particle-based simulation framework. Every particle within distance $d$ from the membrane can react with the Poisson rate $k_{1,m}$.}
	\label{ib5}
\end{figure}
In the particle-based simulation framework we again want to implement this first-order process using a Poisson rate $k_{1,m}$ to come up with a reaction probability $p_{\text{react},1m}(\Delta t)$ that a particle has attached to the membrane within a fixed, finite time step $\Delta t$. Since we want only particles that are in close proximity to the membrane to be able to attach to it, we introduce a finite distance $d$ to encode this spatial confinement (see figure \ref{ib5}). In this way only particles within a distance $z<d$ to the bottom surface at $z=0$ can attempt this first order attachment reaction. In this picture the actual system boundary at $z=0$ is also treated as a reflective no-flux boundary like the opposite boundary at $z=L_z$, since the adsorption step can now happen anywhere in the layer of thickness $d$ above the bottom surface.

To relate the Poisson-rate $k_{1,m}$ to the rate constant $\sigma$ of the mean-field equation \eqref{eq8} we approximate the $\delta$-functional in equation \eqref{eq8} by a $\Theta$-function of width $d$ 
\begin{align}
\delta(z)\ \longrightarrow \ \frac{1}{d} \Theta (d-z) \text{\ \ with\ \ } z > 0,
\end{align}
mimicking our scenario in the particle-based framework. Using $\rho = \rho(z,t)$ and the approximation from above we obtain
\begin{align}
\partial_t\rho= D \partial_{zz}\rho - \frac{\sigma}{d} \Theta (d-z) \rho.
\end{align}
Integrating once over the full domain we get
\begin{align}
\int_{0}^{L_z}  \partial_t\rho\, \mathrm{d}z= \int_{0}^{L_z} \left[ D \partial_{zz}\rho - \frac{\sigma}{d} \Theta (d-z) \rho \right] \, \mathrm{d}z.
\end{align}
The term on the left hand side of the above expression
\begin{align}
\int_{0}^{L_z} \partial_t\rho\, \mathrm{d}z = \partial_t\int_{0}^{L_z} \rho \, \mathrm{d}z  = \partial_t N
\end{align}  is the time derivative of the total amount of particles, while the term
\begin{align}
\int_{0}^{L_z} \Theta (d-z) \rho\, \mathrm{d}z =: N_d
\end{align}
denotes all particles $N_d$ that are less than $d$ away from the membrane at $z=0$.
The term $\int_{0}^{L_z}  D \partial_{zz}\rho\, \mathrm{d}z$ vanishes because of the reflecting boundary conditions ($\partial_z\rho(0) = \partial_z\rho(L_z) = 0$). With this we obtain
\begin{align}
\partial_t N = - \frac{\sigma}{d} N_d.
\end{align}
Since in this setting $N$ can only change by membrane attachment of the particles $N_d$, we can identify $\sigma/d$ as the desired Poisson rate 
\begin{align}
k_{1,m}=\frac{\sigma}{d}
\end{align}
for the first-order reaction step that we wish to implement.
This means that a single particle that is within distance $d$ of a reactive surface of the system has a probability of
\begin{align}
p_{\text{react},1m}(\Delta t) = 1 - \exp(-k_{1,m} \Delta t) = 1- \exp\left(-\frac{\sigma}{d} \Delta t\right)
\end{align}
to react during each time step of length $\Delta t$.

To compare  our simulation algorithm with the mean-field equation \eqref{eq8}, we solved the PDE numerically using an explicit time-stepping scheme for a finite-difference discretization using the Min-system parameter set A as introduced in table 1 of the main text. For this given parameter choice the numerical solutions for the density profiles $\rho(z,t)$ were approximately homogeneous along $z$, due to fast diffusion in comparison to other competing timescales.
Therefore we approximated the solution of equation \eqref{eq8} as a spatially homogeneous density $\rho(z,t) = \rho(t)$ allowing us to obtain an analytical solution for this specific parameter set and system size. Under this assumption equation \eqref{eq8} simplifies to
\begin{align}
\partial_t\rho(t) = - \sigma \delta(z)\rho(t),
\end{align}
which after integrating once over the entire domain gives
\begin{align}
L_z \partial_t \rho(t) = -\sigma \rho(t)
\end{align}
with the standard solution
\begin{align}
\rho(t) = \rho_0 \exp\left(-\frac{\sigma}{L_z} t\right).
\end{align}
\begin{figure}[h!]
	\centering
	\includegraphics[scale=1.0]{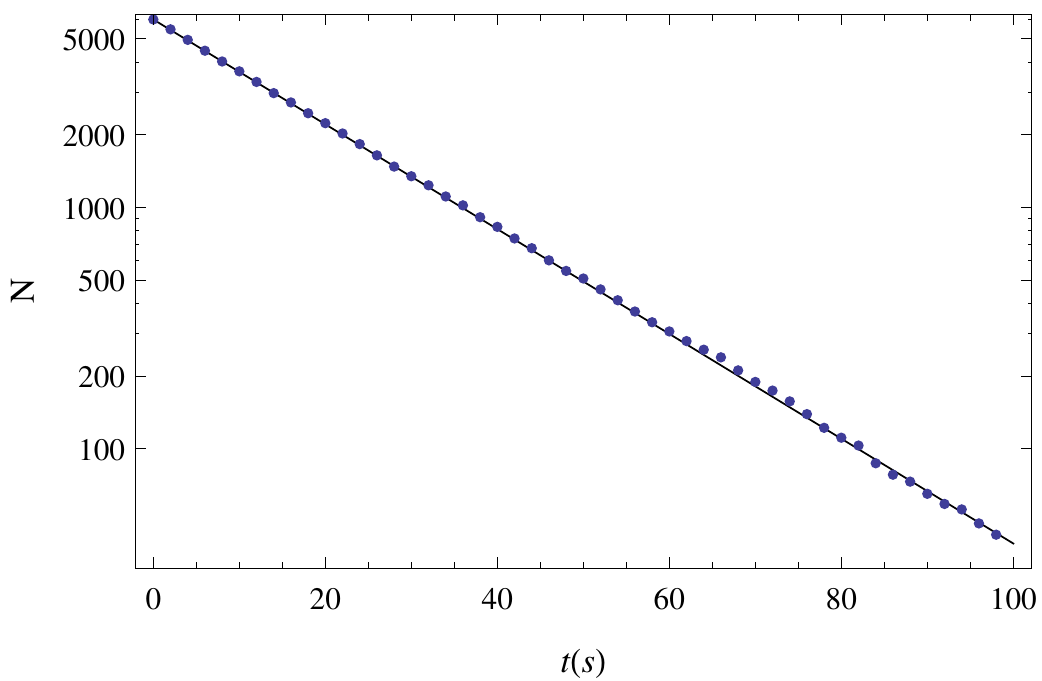}
	\caption{Comparison between the particle-based algorithm (blue points) presented in this section and the mean-field approach (solid line) for the first-order membrane attachment reaction. The analytic solution shown here is based on the assumption of a spatially homogeneous density profile at all times. The y-axis shows the amount of particles $N$ that have not reacted at the membrane yet.}
	\label{ib9}
\end{figure}
Simulation results from the particle-based algorithm are in good agreement with this analytical solution, as depicted in figure \ref{ib9}, where we chose the intrinsic algorithm parameter $d =$\SI{0.02}{\micro\metre}. This choice is also sufficiently small compared to the full domain size of $L_z$ (4\% of $L_z$ for $L_z=\SI{0.5}{\micro\metre}$).

\subsection*{\textbf{$\bm{(iii)}$} Bimolecular membrane attachment}
The third reaction type we consider here describes the bimolecular association of a freely diffusing bulk particle with already membrane-bound particles.
The corresponding mean-field differential equation for this type of reactions reads
\begin{align}
\partial_t\rho = D \nabla^2 \rho - \sigma \delta(z) \tilde{\rho}_\text{b} \rho,
\label{eq9}
\end{align}
where $\rho$ denotes the density of free particles of species $A$ and $\tilde{\rho}_{\text{b}}$
is the density of membrane-bound particles. In the particle-based framework we adopt an algorithm used in the software package Smoldyn \cite{Andrews2004}. The treatment for irreversible bimolecular association reactions follows the spirit of the classical Smoluchowski-model for irreversible diffusion-limited reactions \cite{Smoluchowski1917}.
Here two particles instantaneously react upon collision, leading to the diffusion-limited on-rate for bimolecular association of
\begin{align}
k_S = 4\pi D r_S\, ,
\end{align}
where $D$ denotes the mutual diffusion coefficient of the reacting pair and $r_S$ the effective collision distance which in the most naive picture is taken to be the sum of the molecular radii.
To also address activation-limited or only partially diffusion-influenced reactions Collins and Kimball \cite{Collins1949} proposed to impose a radiation-boundary condition which nicely conveys the picture of a finite activation-limited reaction probability once a pair of reactants has met via diffusional encounter. For the sake of simplicity and algorithmic speed, Andrews and Bray \cite{Andrews2004} came up with an algorithm which keeps the spirit of the original idea by Smoluchowski by introducing a binding radius $\sigma_r$ and letting particles react instantaneously once they are separated by $r<\sigma_b$.
To account for activation-limited effects the sum of the molecular radii is in this picture replaced by this effective binding radius $\sigma_r \le r_S$. They propose an algorithm which calculates $\sigma_r$ for a given diffusion coefficient $D$, time step $\Delta t$ and a forward rate constant $k$. In fact the algorithm solves the forward problem of determining the effective simulated forward rate constant $k$ for a given time step $\Delta t$, diffusion coefficient $D$ and binding radius $\sigma_b$ and stores these results in a look-up table, which is then inverted to interpolate a binding radius for a given reaction rate $k$. The forward problem itself is solved numerically by propagating an initial radial distribution function (RDF) $f(r)$ forward in time, by carrying out alternating diffusion and irreversible reaction steps.
The diffusive steps are implemented by convolving the full radial distribution function with a three-dimensional Gaussian and in each reaction step the RDF is set to zero in the range $0\le r\le \sigma_b$ to account for the irreversible reactions that have taken place. In each iteration the effective reaction rate is then given by integrating the area under the RDF from $[0,\, \sigma_b]$ after the diffusive step. This procedure is repeated iteratively until convergence is achieved. By inverting the tabulated relation between the $k$'s and the $\sigma_b$'s one solves the inverse problem and can thus obtain $\sigma_b = \sigma_b(D,\Delta t, k)$ as a function of the diffusion coefficient $D$, the time step $\Delta t$ and the forward rate constant $k$. The radial distribution functions $f(r)$ one obtains using this algorithm reduce to the radial distribution function of the Smoluchowski model $f_S(r)$ in the limit of small time steps $\Delta t \rightarrow 0$, as to be expected for infinitely detailed Brownian motion. For irreversible bimolecular reactions the RDF in the Smoluchowski model reads
\begin{align}
f_{S}(r)= \begin{cases}
1-\frac{\sigma_b}{r} &\text{for} \quad r\ge \sigma_b \\
0 &\text{for} \quad {r < \sigma_b}.
\end{cases}
\end{align}
In general however, the RDFs qualitatively resemble the functional form of the radial distribution function according to the Collins and Kimball \cite{Collins1949} model.

\begin{figure}[h!]
	\centering
	\includegraphics[scale=1.0]{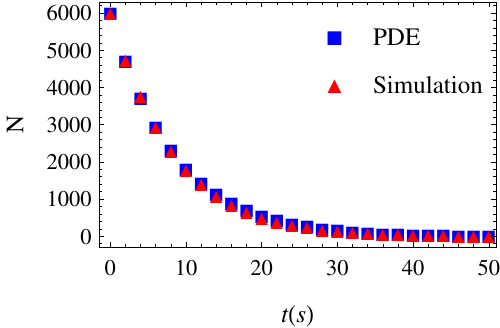}
	\includegraphics[scale=1.0]{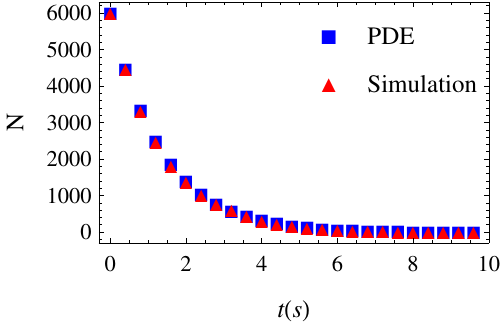}\\
	\includegraphics[scale=1.0]{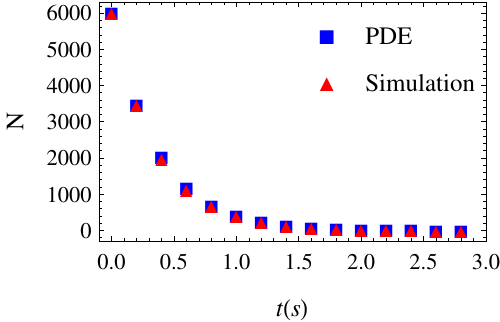}
	\includegraphics[scale=1.0]{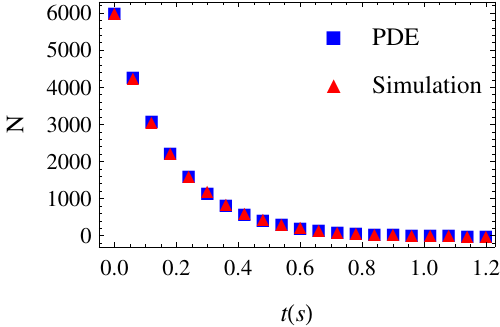}
	\caption{Simulation of $6003$ free particles that are removed upon reaction with evenly distributed bound particles on the bottom surface of a rectangular simulation volume. $N$ denotes number of remaining free particles in the system. Blue rectangles depict the values from the numerical solution of equation \eqref{eq9} and red triangles represent the results from the particle-based simulation algorithm. The parameters from the main text are used. Those are $D=\SI{2.5}{\micro\meter^2/\second}$, $L_z=\SI{0.5}{\micro\metre}$ and $\sigma = \SI{0.0149}{\micro\metre^3/\second}$. In the four figures different values for the density of the bound particles were used: top left $\tilde{\rho}_\text{b}=\SI{4}{\micro\metre^{-2}}$, top right $\tilde{\rho}_\text{b}=\SI{25}{\micro\metre^{-2}}$, bottom left $\tilde{\rho}_\text{b}=\SI{100}{\micro\metre^{-2}}$ and bottom right $\tilde{\rho}_\text{b}=\SI{225}{\micro\metre^{-2}}$.}
	\label{ib10}
\end{figure}

For a quantitative test of our algorithm with the corresponding mean-field results we solve equation \eqref{eq9} numerically. In order to do so we make further simplifying assumptions.
For arbitrary distributions of the bound particles $\tilde{\rho}_\text{b}$ the differential equation \eqref{eq9} can not be reduced to one dimension
as before, however, it is possible to do so for evenly distributed bound particles.
We numerically solved this PDE using the Min-system parameter set A as introduced in table 1 of the main text and assuming evenly distributed bound particles $\tilde{\rho}_\text{b}=\text{const}.$ on the bottom surface of a rectangular simulation box of volume $V$. To match this scenario of the mean-field model in the particle-based framework, we consider a molecular species $A$ of freely diffusing particles in a simulation volume $V$ which can react with membrane-bound particles of another species $B$. For the sake of simplicity we remove particles of species $A$ after a successful reaction step with a membrane-bound particle, while the bound-particles are not removed after the reaction step to keep $\tilde{\rho}_\text{b}=\text{const}$. The results of this comparison using the Min-systems' parameters and several different values for $\tilde{\rho}_{\text{b}}$ are shown in figure \ref{ib10}.

%\bibliographystyle{apsrev4-1}
%\bibliography{references.bib}

%merlin.mbs apsrev4-1.bst 2010-07-25 4.21a (PWD, AO, DPC) hacked
%Control: key (0)
%Control: author (72) initials jnrlst
%Control: editor formatted (1) identically to author
%Control: production of article title (-1) disabled
%Control: page (0) single
%Control: year (1) truncated
%Control: production of eprint (0) enabled
%

\end{document}